\documentclass{aa}
\usepackage{graphicx}
\usepackage{txfonts}
\usepackage[]{natbib}
\usepackage[]{longtable}
\usepackage{multirow}

\bibpunct{(}{)}{;}{a}{}{,}

\begin{document}
   \title{Rise and fall of the X-ray flash 080330: an off-axis jet?}


   \author{C.~Guidorzi\inst{1,2,3}, C.~Clemens\inst{4}, S.~Kobayashi\inst{2}, J.~Granot\inst{5},
A.~Melandri\inst{2}, P.~D'Avanzo\inst{1}, N.~P.~M.~Kuin\inst{6}, A.~Klotz\inst{7,8}, J.~P.~U. Fynbo\inst{9},
S.~Covino\inst{1}, J.~Greiner\inst{4}, D.~Malesani\inst{9}, J.~Mao\inst{1,10}, C.~G.~Mundell\inst{2},
I.~A.~Steele\inst{2}, P.~Jakobsson\inst{11}, R.~Margutti\inst{12,1}, D.~Bersier\inst{2}, S.~Campana\inst{1},
G.~Chincarini\inst{12,1}, V.~D'Elia\inst{13}, D.~Fugazza\inst{1}, F.~Genet\inst{5}, A.~Gomboc\inst{14},
T.~Kr\"uhler\inst{4}, A.~K\"upc\"u~Yolda\c s\inst{4}, A.~Moretti\inst{1},
C.~J.~Mottram\inst{2}, P.~T.~O'Brien\inst{15}, R.~J.~Smith\inst{2}, G.~Szokoly\inst{4},
G.~Tagliaferri\inst{1}, N.~R.~Tanvir\inst{15}, N.~Gehrels\inst{16}
}

   \offprints{C.\ Guidorzi, guidorzi@fe.infn.it}
   \institute{INAF -- Osservatorio Astronomico di Brera, via E. Bianchi 46, I-23807 Merate (LC), Italy
   \and Astrophysics Research Institute, Liverpool John Moores University, Twelve Quays House,
        Birkenhead, CH41 1LD, UK
   \and Dipartimento di Fisica, Universit\`a di Ferrara, via Saragat 1, I-44100 Ferrara, Italy
   \and Max-Planck-Institut f\"ur extraterrestrische Physik, 85740 Garching, Germany
   \and Centre for Astrophysics Research, University of Hertfordshire, College Lane, Hatfield, Herts AL10 9AB, UK
   \and Mullard Space Science Laboratory/UCL, Holmbury St Mary, Dorking, Surrey RH5 6NT, UK
   \and Observatoire de Haute-Provence, 04870 Saint Michel l'Observatoire, France
   \and CESR, Observatoire Midi-Pyr\'en\'ees, CNRS, Universit\'e de Toulouse, BP 4346, 31028 Toulouse,
        Cedex 04, France
   \and Dark Cosmology Centre, Niels Bohr Institute, University of Copenhagen, Juliane Maries
        vej 30, DK-2100 K\o{}benhavn \O , Denmark
   \and Yunnan  Observatory,  National Astronomical Observatories, Chinese Academy of Sciences, P.O.~Box 110,
        Kunming, Yunnan Province, 650011, China
   \and Centre for Astrophysics and Cosmology, Science Institute, University of Iceland, Dunhagi 5,
        IS 107 Reykjavik, Iceland
   \and Dipartimento di Fisica, Universit\`a di Milano-Bicocca, piazza delle Scienze 3, 
        I-20126 Milano, Italy
   \and INAF -- Osservatorio Astronomico di Roma, via Frascati 33, I-00040 Monteporzio (RM), Italy
   \and Faculty of Mathematics and Physics, University of Ljubljana, Jadranska 19, SI-1000 Ljubljana, Slovenia
   \and Department of Physics and Astronomy, University of Leicester, Leicester, LE1 7RH, UK
   \and NASA Goddard Space Flight Center, Greenbelt, MD 20771, USA
             }

   \date{\today}
%
\abstract
{X-ray flashes (XRFs) are a class of gamma-ray bursts (GRBs) with the peak energy of the time-integrated
$\nu\,F_\nu$ spectrum, $E_{\rm p}$, typically below 30~keV, whereas classical GRBs have $E_{\rm p}$ of
a few hundreds keV. Apart from $E_{\rm p}$ and the systematically lower luminosity, the properties of XRFs,
such as the duration or the spectral indices, are typical of the classical GRBs.
Yet, the nature of XRFs and the differences from that of GRBs are not understood.
In addition, there is no consensus on the interpretation of the shallow decay phase observed in most
X-ray afterglows of both XRFs and GRBs.}
{We examine in detail the case of XRF~080330 discovered by Swift at the redshift of $1.51$. This burst is
representative of the XRF class and exhibits an X-ray shallow decay. The rich and broadband
(from NIR to UV) photometric data set we collected across this phase makes it an ideal candidate to test the
off-axis jet interpretation proposed to explain both the softness of XRFs and the shallow decay phase.}
{We present prompt $\gamma$-ray, early and late NIR/visible/UV and X-ray observations of the
XRF~080330. We derive a spectral energy distribution from NIR to
X-ray bands across the shallow/plateau phase and we describe the temporal evolution
of the multi-wavelength afterglow within the context of the standard afterglow model.}
{The multi-wavelength evolution of the afterglow is achromatic from $\sim$$10^2$~s out to
$\sim$$8\times10^{4}$~s. The energy spectrum from NIR to X-ray is nicely fitted with a simple power-law,
$F_\nu\propto\nu^{-\beta_{\rm ox}}$, with $\beta_{\rm ox}=0.79\pm0.01$ and negligible rest-frame dust extinction.
The light curve can be modelled either by a piecewise power-law or by the combination of a smoothly
broken power law with an initial rise up to $\sim$600~s, a plateau lasting up to $\sim$2~ks,
followed by a gradual steepening to a power-law decay index of $\sim$$2$ out to 82~ks.
At this point, there appears a bump modelled with a second component, while the corresponding
optical energy spectrum, $F_\nu\propto\nu^{-\beta_{\rm o}}$, reddens by $\Delta\beta_{\rm o}=0.26\pm0.06$.}
{A single-component jet viewed off-axis explains the light curve of XRF~080330,
the late time reddening as due to the reverse shock of an energy injection episode, and its being an XRF.
Other possibilities, such as the optical rise marking the pre-deceleration of the fireball
within a wind environment, cannot be definitely excluded, but seem somewhat contrived.
We rule out the dust decreasing column density swept up by the fireball as the explanation of
the rise of the afterglow.}

{}
\authorrunning {C.\ Guidorzi  et al.}
\titlerunning {Prompt and afterglow study of XRF~080330}
\maketitle

\section{Introduction}
\label{sec:intro}
Time-integrated photon spectra of long gamma-ray bursts (GRBs) can be
adequately fitted with a smoothly broken power law \citep{Band93}, whose
low-energy and high-energy indices, $\alpha_{\rm B}$ and $\beta_{\rm B}$,
have median values of $-1$ and $-2.3$, respectively \citep{Preece00,Kaneko06}.
The corresponding $\nu$$F_\nu$ spectrum peaks at $E_{\rm p}$,
the so-called peak energy, whose rest frame value is found to correlate
with other relevant observed intrinsic properties, such as the
isotropic-equivalent radiated $\gamma$-ray energy, $E_{\rm iso}$
\citep{Amati02},
or its collimation-corrected value, $E_{\gamma}$ \citep{Ghirlanda04}. 
In the BATSE catalogue, the $E_{\rm p}$ distribution clusters around
$300$~keV with a $\sim$$100$~keV width \citep{Kaneko06}.

When observations of GRBs in softer energy bands than BATSE became
available thanks to BeppoSAX and HETE-2, a new class of soft GRBs with
$E_{\rm p}$$\lesssim30$~keV, so named X-ray flashes (XRFs),
was soon discovered \citep{Heise01,Barraud03}.
GRBs with intermediate softness, called X-ray rich (XRR)
bursts, were also observed, with $30$~keV~$\la$~$E_{\rm p}$~$\la$$100$~keV
\citep{Sakamoto05}.
These soft GRBs share the same temporal and spectral properties, aside
from the systematically lower $E_{\rm p}$, with the classical GRBs both
for the prompt \citep{Frontera00,Barraud03,Amati04} and, partially, the afterglow
emission \citep{Sakamoto05,Dalessio06,Mangano07}.
Moreover, they were found to obey the $E_{\rm p}$--$E_{\rm iso}$ correlation
\citep{Amati07} discovered for classical GRBs, extending it all the way down
to $E_{\rm p}$ values of a few keV and forming a continuum \citep{Sakamoto05}.
Like classical GRBs, also XRFs have been found to be associated with SNe
\citep{Campana06,Pian06} and therefore connected with the collapse of massive stars.
The comparison also holds for the cases in which apparently no associated SN was found both
for classical GRBs (e.g. Della~Valle et~al. 2006, Fynbo et~al. 2006, Gal-Yam et~al. 2006)
\nocite{DellaValle06,Fynbo06,Galyam06} and for XRFs \citep{Levan05}.

A number of different models have been proposed in the literature to explain
the nature of XRRs and XRFs (e.g., see the review of Zhang 2007):\nocite{Zhang07_rev}
i) standard GRBs viewed well off the axis of the jet, thus explaining the
softness as due to a larger viewing angle and a lower Doppler factor
\citep{Yamazaki02,Granot02,Granot05_2}; ii) two coaxial jets with different opening angles
(wide and narrow), $\theta_{\rm w}>\theta_{\rm n}$, and viewed at an angle
$\theta_{\rm v}$, $\theta_{\rm n}<\theta_{\rm v}<\theta_{\rm w}$ \citep{Peng05};
iii) the ``dirty fireball'' model characterised
by a small value of the bulk Lorentz factor due to a relatively high baryon loading
of the fireball \citep{Dermer00}; iv)  distribution of high Lorentz factors with
low contrast of the colliding shells \citep{Mochkovitch04}.
In the off-axis interpretation, a number of different models of the structure
and opening angle of the jet have been proposed (e.g. Granot et~al. 2005; Donaghy 2006).
\nocite{Donaghy06,Granot05_2}

The advent of Swift \citep{Gehrels04} has made it possible to collect a large sample of
early X-ray afterglow light curves of GRBs. Concerning the XRRs and XRFs, the 15--150~keV
energy band of the Swift Burst Alert Telescope (BAT; Barthelmy et~al. 2005)
\nocite{Barthelmy05} and its relatively large effective area still allow
to detect them, although those with $E_{\rm p}$ of a few keV are disfavoured
with respect to BeppoSAX and HETE-2 instruments \citep{Sakamoto08}.
Thanks to Swift it is possible to study the early X-ray afterglow properties
of these soft events. Like hard GRBs, also XRFs occasionally exhibit
X-ray flares \citep{Romano06}. \citet{Sakamoto08} analysed a sample
of XRFs, XRRs and classical GRBs detected with Swift and found some evidence
for an average X-ray afterglow luminosity of XRFs being roughly half that
of classical GRBs and some differences between the average X-ray afterglow light curves.

An unexpected discovery of Swift is the shallow decay phase experienced by
most of X-ray afterglows between a few $10^2$ up to $10^3$--$10^4$~s after the
trigger time \citep{Tagliaferri05,Nousek06,Zhang06}.
Several interpretations have been put forward (e.g. see Ghisellini et~al.
2009 for a brief review). \nocite{Ghisellini09}
Among them, some invoke continuous energy injection to the fireball shock front
through refreshed shocks \citep{Nousek06,Zhang06}, depending on the progenitor period
of activity: a long- or short-lived powering mechanism, either in the form of a
prolonged, continuous energy release ($L(t)\propto$$t^{-q}$), or via discrete 
shells whose $\Gamma$ distribution is a steep power-law.
For instance, in the cases of GRB~050801 and GRB~070110 a newly born millisecond
magnetar was suggested to power the flat decay observed in the optical and X-ray bands
\citep{Depasquale07, Troja07}.
Alternatively, geometrical models interpret the shallow decay as the delayed onset
of the afterglow observed from viewing angles outside the edge of a jet
\citep{Granot02,Salmonson03,Granot05,Eichler06}.
Other models invoke two-component jets viewed
off the axis of the narrow component, 
also invoked to explain the late time observations of GRB~030329 \citep{Berger03}.
In particular, this model would explain the initial flat decay observed
in XRF~030723, dominated by the wide component, followed by a late rebrightening
peaking at $\sim$16 days and interpreted as due to the deceleration and lateral expansion
of the narrow component \citep{Huang04,Butler05}, although alternative explanations
for this in terms of a SN have also been proposed \citep{Fynbo04,Tominaga04}.

Other models explain the shallow decay as due to a temporal evolution of the
fireball micro-physical parameters \citep{Ioka06,Granot06}; scattering by dust located
in the circumburst medium \citep{Shao07}; ``late prompt'' activity of the inner
engine, keeping up a prolonged emission of progressively lower power and Lorentz
factor shells, which radiate at the same distance as for the prompt emission
\citep{Ghisellini07,Ghisellini09}; a dominating reverse shock in the X-ray
band propagating through late shells with small Lorentz factors \citep{Genet07,Uhm07}.
\citet{Yamazaki09} suggests that the plateau and the following standard decay
phases are an artifact of the choice of $t_0$, provided that the engine activity
begins before the trigger time by $\sim$$10^3$--$10^4$~s.

XRF~080330 was promptly discovered by the Swift-BAT and automatically pointed with
the X-Ray Telescope (XRT; Burrows et~al. 2005)\nocite{Burrows05} and the
Ultraviolet/Optical Telescope (UVOT; Roming et~al. 2005)\nocite{Roming05} as shown
in Figure~\ref{f:multi_lc}.
In this work we present a detailed analysis of the Swift data, from the prompt
$\gamma$-ray emission to the X-ray and optical afterglow and combine it with
the large multi-filter data set collected from the ground, encompassing a broad band,
from NIR to UV wavelengths, and spanning from one minute out to
$\sim$3 days post burst. The main properties exhibited by XRF~080330 are the
rise of the optical afterglow up to $\sim$300~s, followed by a shallow decay
also present in the X-ray, after which it gradually steepens, and either a possible
late time ($\sim10^5$~s) brightening (Fig.~\ref{f:fluxall_beulore}) or a sharp
break (Fig.~\ref{f:fluxall_beu3}).
The richness of the multi-wavelength data collected throughout the rise-flat top-steep
decay allows us to constrain the broadband energy spectrum of the shallow decay
phase as well as its spectral evolution. Moreover, it is possible to constrain
the optical flux extinction due to dust along the line of sight and, in particular,
near the progenitor.
This GRB is a good benchmark for the proposed models of XRFs sources
and of their link with the classical GRBs through the common properties, such
as the flat decay phase.

The paper is organised as follows: Sects.~\ref{sec:obs} and \ref{sec:an} report
the observations, data reduction and analysis, respectively.
We report our multi-wavelength combined analysis in Sect.~\ref{sec:multi}.
In Sect.~\ref{sec:disc}
we discuss our results in the light of the models proposed in the literature
and Sect.~\ref{sec:conc} reports our conclusions.

Throughout the paper, times are given relative to the BAT trigger time.
The convention $F(\nu,t)\propto\nu^{-\beta}\,t^{-\alpha}$
is followed, where the spectral index $\beta$ is related to the
photon index $\Gamma=\beta+1$.
We adopted the standard cosmology: $H_0=70$\,km\,s$^{-1}$\,Mpc$^{-1}$,
$\Omega_\Lambda=0.7$, $\Omega_{\rm M}=0.3$.

All the quoted errors are given at 90\% confidence level for one interesting
parameter ($\Delta\chi^2=2.706$), unless stated otherwise.

\section{Observations}
\label{sec:obs}
XRF~080330 triggered the {\rm Swift}-BAT on 2008 March 30 at 03:41:16~UT.
The $\gamma$-ray prompt emission in the 15--150~keV energy band consisted of a multiple--peak
structure with a duration of about 60~s \citep{Mao08a}.
An uncatalogued, bright and fading X-ray source was promptly identified by XRT.
From the initial 100-s finding chart taken with the UVOT telescope in the White filter
from 82~s the optical counterpart was initially localised at RA $= 11^{\rm h}$  $17^{\rm m}$  $04\fs51$,   
Dec. $=+30^{\circ}$ $37^{\prime}$ $22\farcs1$ (J2000), with an error radius of $1\farcs0$ (1$\sigma$;
Mao et~al. 2008a).\nocite{Mao08a}
During the observations, the Swift star trackers failed to maintain a proper
lock resulting in a drift which affected the observations and accuracy of early reports.
We finally refined the position from the UVOT field match to the USNO--B1 catalogue:
RA $= 11^{\rm h}$  $17^{\rm m}$  $04\fs52$, Dec. $=+30^{\circ}$ $37^{\prime}$ $23\farcs5$ (J2000),
with an error radius of $0\farcs3$ (1$\sigma$; Mao et~al. 2008b),\nocite{Mao08b}
consistent with the position derived from ground telescopes (e.g., PAIRITEL, Bloom \& Starr 2008).
\nocite{Bloom08}

The T\'elescopes \`a Action Rapide pour les Objets Transitoires (TAROT; Klotz et~al. 2008c)\nocite{Klotz08c}
began observing at $20.4$~s ($4.5$~s after the notice) and discovered independently
the optical counterpart during the rise with $R$$\sim$$16.8$ at 300~s \citep{Klotz08a}.
TAROT went on observing until the dawn at $1.4$~ks \citep{Klotz08b}.

The Rapid Eye Mount\footnote{{\tt http://www.rem.inaf.it/}} (REM; Zerbi et al. 2001)\nocite{Zerbi01}
telescope reacted promptly and began observing at 55~s and detected the optical afterglow
in $R$ band \citep{Davanzo08}.
The optical counterpart was promptly detected also by other robotic telescopes, such as 
ROTSE--IIIb \citep{Schaefer08,Yuan08}, PROMPT \citep{Schubel08} and RAPTOR; the latter in particular observed
a $\sim$$10$-s long optical flash of $R=17.46\pm0.22$ at 60~s contemporaneous with the last
$\gamma$-ray pulse \citep{Wren08}.

The Liverpool Telescope (LT) began observing at 181~s. The optical afterglow was automatically identified
by the LT-TRAP GRB pipeline \citep{Guidorzi06} with $r'$$\sim$$17.3$ \citep{Gomboc08a}, thus triggering the
multi-colour imaging observing mode in the $g'r'i'$ filters which lasted up to the dawn at $4.9$~ks.

The Faulkes Telescope North (FTN) observations of XRF~080330 were carried out from $8.4$ to $9.1$~hr and
again from $31.8$ to $33.9$~hr with deep $r'$ and $i'$ filter exposures.

The Gamma-Ray Burst Optical and Near-Infrared Detector (GROND; Greiner et~al. 2008)\nocite{Greiner08}
started simultaneous observations in $g'r'i'z'JHK$ filters of the field of GRB~080330 at $3.1$~minutes
and detected the afterglow with $J=15.92\pm0.04$ and $H=15.46\pm0.11$ from the first 240-s of effective
exposure \citep{Clemens08}.

A spectrum of XRF~080330 was acquired at $46$~minutes with the Nordic Optical Telescope (NOT).
The identification of absorption features allowed to measure the redshift, which turned out to
be $z=1.51$ \citep{Malesani08}. This was soon confirmed by the spectra taken with the
Hobby-Eberly Telescope \citep{Cucchiara08}.

The Galactic reddening along the line of sight to the GRB is $E_{B-V}=0.017$ \citep{Schlegel98}.
The corresponding extinction in each filter was estimated through the NASA/IPAC Extragalactic Database
extinction calculator\footnote{{\tt http://nedwww.ipac.caltech.edu/forms/calculator.html}}:
$A_{UVW1}=0.120$, $A_U=0.090$, $A_B=0.071$, $A_g=0.064$, $A_V=0.055$, $A_r=0.047$, $A_R=0.044$,
$A_I=0.032$, $A_i=0.035$, $A_z=0.022$, $A_J=0.015$, $A_H=0.010$, $A_K=0.006$.

\section{Data reduction and analysis}
\label{sec:an}

\subsection{Gamma--ray data}
\label{sec:gamma}

The BAT data were processed with the {\tt heasoft} package (v.6.4) adopting the ground-refined
coordinates provided by the BAT team \citep{Markwardt08}.
The BAT detector quality map was obtained by processing
the nearest-in-time enable/disable map of the detectors.

\begin{figure}
\centering
\includegraphics[width=8.5cm]{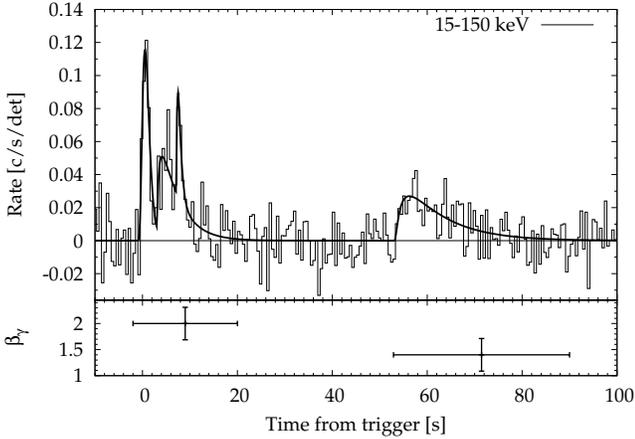}
\caption{{\em Top panel}: 15--150~keV BAT mask-weighted light curve
(binning time of 0.512~s). The thick solid line shows the result
of fitting the profile with four pulses modelled with Norris profiles
\citep{Norris05}.
{\em Bottom panel}: spectral index $\beta_\gamma$ as a function of time.
}
\label{f:bat_lc}
\end{figure}

The top panel of Figure~\ref{f:bat_lc} shows the 15--150~keV
mask-weighted light curve of XRF~080330 as recorded by BAT, expressed
as counts per second per fully illuminated detector for an equivalent on-axis source.
The solid line displayed in Fig.~\ref{f:bat_lc} corresponds to the result of fitting
the profile from $-1.2$ to $100$~s with a combination of four pulses \citep{Markwardt08} as modelled
by Norris et~al. (2005; hereafter N05 model). \nocite{Norris05}
Table~\ref{t:BAT_N05} reports the corresponding derived parameters:
$t_{\rm p}$ (peak time), $A$ (15--150 keV peak flux), $\tau_{\rm r}$ (rise time),
$\tau_{\rm d}$ (decay time), $w$ (pulse width), $k$ (pulse asymmetry) and the model
fluence in the 15--150 keV band. The goodness of the fit is $\chi^2/{\rm dof}=375/379$.
Parameter uncertainties were derived by propagation starting from the
best-fit parameters and taking into account their covariance.
We tried to apply the same analysis to the light curves of the resolved energy channels
to investigate temporal lags and, more generally, the dependence of the parameters on energy;
however, because of the faintness and softness of the signal, we could not
constrain the parameters in a useful way.

\begin{table*}
 \begin{center}
 \caption{Best-fit parameters of the 15--150~keV profile decomposed into four pulses using the
model by \citet{Norris05}. Uncertainties are 1$\sigma$.}
 \label{t:BAT_N05}
 \begin{tabular}{lccccccc}
 \hline
 \hline
 \noalign{\smallskip}
 Pulse  & $t_{\rm p}$ & $A$     & $\tau_{\rm r}$  & $\tau_{\rm d}$  & $w$   & $k$ & Fluence\\
  & (s) & ($10^{-8}$~erg~cm$^{-2}$~s$^{-1}$) &  (s)  & (s) & (s)  &  & ($10^{-8}$~erg~cm$^{-2}$)\\
\noalign{\smallskip}
\hline
\noalign{\smallskip}
    1    &  $0.5\pm0.2$ & $4.6\pm0.6$ & $0.79\pm0.18$ & $1.30\pm0.26$ & $2.09\pm0.29$ & $0.24\pm0.16$ & $8.6\pm1.0$  \\
    2    &  $4.2\pm0.6$ & $2.0\pm0.3$ & $1.04\pm0.64$ & $4.71\pm1.30$ & $5.75\pm1.44$ & $0.64\pm0.20$ & $10.8\pm1.9$ \\
    3    &  $7.5\pm0.2$ & $2.5\pm0.7$ & $0.32\pm0.19$ & $0.84\pm0.43$ & $1.16\pm0.46$ & $0.45\pm0.32$ & $2.7\pm0.8$  \\
    4    & $56.2\pm1.2$ & $1.3\pm0.2$ & $2.52\pm1.26$ & $10.5\pm2.3$  & $13.0\pm2.4$  & $0.61\pm0.18$ & $15.5\pm2.3$ \\
\noalign{\smallskip}
 \hline
 \end{tabular}
 \end{center} 
\end{table*}

In addition, the energy spectra in the 15--150~keV band were extracted using the tool {\tt batbinevt}.
We applied all the required corrections: we updated them through {\tt batupdatephakw} and generated
the detector response matrices using {\tt batdrmgen}. Then we used {\tt batphasyserr} in order to account
for the BAT systematics as a function of energy.
Finally we grouped the energy channels of the spectra by imposing a 3$\sigma$ (or 2-$\sigma$ when the S/N
was too low) threshold on each grouped channel.
We fitted the resulting photon spectra, $\Phi(E)$ (ph cm$^{-2}$s$^{-1}$keV$^{-1}$),
with a power law with pegged normalisation ({\tt pegpwrlw} model under {\tt xspec} v.11.3.2).
We extracted several spectra in different time intervals: over $T_{90}$, total, spanning the bunch
of the first three pulses, the fourth pulse alone and that around the peak, determined on a minimum
significance criterion. The results are reported in Table~\ref{t:BAT_XRT_spec}.
The time-averaged spectral index is $\beta_\gamma=1.65\pm0.51$ with a total fluence of
$S(15-150~{\rm keV})=(3.6\pm0.8)\times10^{-7}$~erg~cm$^{-2}$ and a $0.448$-s peak photon flux of
$(1.0\pm0.2)$~ph~cm$^{-2}$~s$^{-1}$, in agreement with previous results \citep{Markwardt08}.
The bottom panel of Fig.~\ref{f:bat_lc} shows marginal evidence for a soft--to--hard evolution:
$\beta_\gamma$ passes from $2.0\pm0.5$ (first three pulses) to $1.4\pm0.5$ (fourth pulse).

Following \citet{Sakamoto08}, a GRB is classified as an XRR (XRF) depending on
whether the fluence ratio $S(25-50~{\rm keV})/S(50-100~{\rm keV})$ is lower
(greater) than $1.32$. The fluence ratio of XRF~080330, $1.5_{-0.3}^{+0.7}$, places it among the
XRFs, although still compatible with being an XRR burst.
Although from BAT data alone we could not measure the peak energy of the time-integrated
$\nu\,F_\nu$ spectrum, we tried to fit with a smoothed broken power law model \citep{Band93}
by fixing the low-energy index $\alpha_{\rm B}$ to -1 \citep{Kaneko06},
given that $\Gamma=\beta_{\gamma}+1>2$ and is very likely dominated by the high-energy
index $\beta_{\rm B}$.
This way we derived the following constraint: $E_{\rm p}<35$~keV, in agreement with
the upper limit to its rest-frame (intrinsic) value, $E_{\rm p,i}=E_{\rm p}\,(1+z)<88$~keV,
obtained by \citet{Rossi08}.

Recently, \citet{Sakamoto09} calibrated a method aimed at estimating $E_{\rm p}$ from
the $\Gamma_\gamma$ as measured with BAT, provided that $1.3$$<$$\Gamma$$<$$2.3$.
In the case of XRF~080330, the confidence interval on $\Gamma$, $2.65\pm0.51$, marginally
overlaps with the allowed range; however, since $E_{\rm p}$ is anti-correlated with $\Gamma$
and the lower limit on $\Gamma$ lies within the usable range, we can derive an upper limit
to $E_{\rm p}$ from the relation of \citet{Sakamoto09}, which turns out to be $30$~keV,
in agreement with our previous value.
These results are fully consistent with a previous preliminary analysis \citep{Markwardt08}.

We constrained $E_{\rm iso}$ in the rest-frame 1--$10^4$~keV band using the
upper limit on $E_{\rm p}$ of $35$~keV.
Following the prescriptions by \citet{Amati02} and \citet{Ghirlanda04}, we found
$E_{\rm iso}<2.2\times10^{52}$~ergs.
Combined with $E_{\rm p,i}<88$~keV, this places XRF~080330 in the $E_{\rm p,i}$--$E_{\rm iso}$
space consistently with the Amati relation \citep{Amati02,Amati06}.

\subsection{X--ray data}
\label{sec:X}
The XRT data were processed using the {\tt heasoft} package (v.6.4).
We ran the task {\tt xrtpipeline} (v.0.11.6) applying calibration and
standard filtering and screening criteria. Data from 77 to 134~s were
acquired in Windowed Timing (WT) mode and the following in Photon
Counting (PC) mode due to the faintness of the source.  Events with
grades 0--2 and 0--12 were selected for the two modes, respectively.
XRT observations went on up to $5.9\times10^5$~s, with a total net
exposure time of $30.6$~ks.  The XRT analysis was performed in the
0.3--10 keV energy band.

Source photons were extracted from WT mode data in a rectangular region
40 pixels along the image strip (20 pixel wide) centred on the source,
whereas the background photons were extracted from an equally-sized region with
no sources.

Firstly, we extracted the 
first orbit PC data from 136 to 331~s, where the point spread function (PSF)
of the source looked unaffected by the spacecraft drifting, and extracted the following
refined position: RA$= 11^{\rm h}$  $17^{\rm m}$  $04\fs68$,
Dec. $=+30^{\circ}$ $37^{\prime}$ $24\farcs8$ (J2000), with an error radius of
$4.0$~arcsec \citep{MaoGuidorzi08}.
We corrected these data for pile-up by extracting source photons from
an annular region centred on the above position and with inner and outer
radii of 4 and 30 pixels (1~pixel$\mbox{}=2\farcs36$), respectively.
The background was estimated from a three-circle region with a total area of
$30.3\times10^3$~pixel$^{2}$ away from any source present in the field.
Finally, we re-extracted the source photons over the entire first orbit within
a larger circular region centred on the same position and with a radius of 40 pixels,
to compensate for the drifting. The light curve of the full first orbit data
(PC mode) was then corrected so as to match the previous one
correctly produced in the 136--331~s sub-interval.

The resulting 0.3--10~keV light curve is shown in Fig.~\ref{f:multi_lc} (black empty
triangles).
It was binned so as to achieve a minimum signal to noise ratio (SNR) of 3.
The data taken in following orbits were not enough to provide a significant detection
and only a 3$\sigma$ upper limit was obtained.

The X-ray curve can be fitted with a broken power law, with the following
parameters: $\alpha_{\rm x,1}=4.8\pm0.4$, $t_{\rm b}=163_{-10}^{+9}$~s,
$\alpha_{\rm x,2}=0.26\pm0.10$ ($\chi^2/{\rm dof}=66/70$).
The last upper limit clearly requires a further break. We set a lower limit
on $\alpha_{\rm x,3}$ by connecting the end of first orbit data with the late upper limit
under the assumption that the second break occurred at the beginning of the
data gap. This turned into $\alpha_{\rm x,3}>1.3$ ($\ge$3$\sigma$ confidence).
The later the second break time, the steeper the final decay.

We extracted the 0.3--10~keV spectrum in two different time intervals: i) ``XRT-WT''
interval, from $77$ to $134$~s (WT mode), corresponding to the initial steep decay;
ii) ``plateau'' interval, from $423$ to $1507$~s (PC mode) corresponding to the
following flat decay (or ``plateau'') phase.
Source and background spectra were extracted from the same regions as the ones
used for the light curve for the corresponding time intervals and modes.
The ancillary response files were generated using the task {\tt xrtmkarf}.
Spectral channels were grouped so as to have at least 20 counts per bin.
Spectral fitting was performed with {\tt xspec} (v.~11.3.2).

We modelled both spectra with a photoelectrically absorbed power law
(model {\tt wabs$\cdot$zwabs$\cdot$pow}), adopting the photoelectric cross section by \citet{Morrison83}.
The first column density was frozen to the weighted average Galactic value
along the line of sight to the GRB, $N_{\rm H}^{\rm (Gal)}=1.23\times10^{20}$~cm$^{-2}$
\citep{Kalberla05}, while the second rest-frame column density, $N_{{\rm H},z}$, was
left free to vary. While during the steep decay we found no evidence for significant rest-frame
absorption, with a 90\% confidence limit of $N_{{\rm H},z}<1.4\times10^{21}$~cm$^{-2}$,
in the plateau spectrum we found only marginal evidence for it,
$N_{{\rm H},z}=1.6_{-1.5}^{+1.8}\times10^{21}$~cm$^{-2}$. The spectral index, $\beta_{\rm x}$,
varies from $1.06_{-0.09}^{+0.10}$ to $0.80_{-0.15}^{+0.16}$: the significance of this change
is $\sim2.3$~$\sigma$.
The best-fit parameters are reported in Table~\ref{t:BAT_XRT_spec}.

\subsection{Near--UV/Visible UVOT data}
\label{sec:UVOT}
The Swift UVOT instrument started observing on 2008 March 30 at 03:42:19~UT,
63~s after the BAT trigger, with a $9.37$-s settling
exposure. Since the detectors are powered up during this exposure, the
effective exposure time may be less than reported. We checked the
brightness in this exposure with later exposures, to confirm that no
correction was needed. The first $99.7$-s finding chart exposure
started at 03:42:39~UT in the white filter in event mode followed by
a $399.8$-s exposure in the $V$ filter, also in event mode.  Due to
the loss of lock by the spacecraft star trackers, the attitude
information was incorrect. In order to process the data, {\tt xselect}
was used to extract images for short time intervals. The length of the
interval was chosen short enough that the drift of the spacecraft was
mostly within $7\arcsec$, and at most $14\arcsec$, but long enough to
get a reasonably accurate measurement.
A source region was placed over the position of
the source, making checks for consistency with the position of nearby
stars, and the magnitudes were determined using the ftool {\tt uvotevtlc}.
In most cases, an aperture with a radius of $5\arcsec$ was used, and three
measurements used a slightly larger aperture. No aperture correction
was made, since the source shape in those cases was very elongated.
The measured magnitudes were converted back to the original count
rates, using the UVOT calibration \citep{Poole08}.  These were
subsequently converted to fluxes using the method of \citet{Poole08},
but for an incident power-law spectrum with $\beta$$=$$0.8$ and for
a redshift $z$$=$$1.51$.

\subsection{NIR/Visible ground-based data}
\label{sec:opt}

Robotically triggered observations with the LT began at 181~s leading to the automatic
identification by the GRB pipeline LT-TRAP \citep{Guidorzi06} of the optical afterglow
at the position RA $= 11^{\rm h}$ $17^{\rm m}$ $04\fs48$, Dec. $=+30^{\circ}$ $37^{\prime}$ $23\farcs8$
(J2000; 1$\sigma$ error radius of $0.2$~arcsec).
This is consistent within $1.6$~$\sigma$ with the refined UVOT position.
The afterglow was seen during the end of the rise with $r'=17.3\pm0.1$
and subsequently decay \citep{Gomboc08a}.
Following two initial sequences of $3\times10$~s each in the $r'$ filter during the detection
mode (DM), the multi-colour imaging observing mode (MCIM) in the $g'r'i'$ filters was
automatically selected. Observations carried on up to $4.9$~ks.

The FTN observations of XRF~080330 were carried out from $8.4$ to $9.1$~hr and again from
$31.8$ to $33.9$~hr with deep $r'$ and $i'$ filter exposures as part of the {\em RoboNet~1.0}
project\footnote{{\tt http://www.astro.livjm.ac.uk/RoboNet/}} \citep{Gomboc06}.

Calibration was performed against five non-saturated field stars with preburst SDSS
photometry \citep{Cool08}, by adopting their PSF to adjust the zero point of the single
images. Photometry was carried out using the Starlink GAIA software.
Magnitudes were converted into flux densities (mJy) following \citet{Fukugita96}.
Results are reported in Table~\ref{tab:photom}.

Optical $R$-band observations of the afterglow of GRB~080330 were carried out with the REM
telescope equipped with the ROSS optical spectrograph/imager on 2008 March 30, starting about~55
seconds after the burst \citep{Davanzo08}.
We collected 38 images with typical exposures times of 30, 60 and 120~s, covering a
time interval of about $0.5$~hours.
Image reduction was carried out by following the standard procedures:
subtraction of an averaged bias frame, division by a normalised flat frame.
The astrometry was fitted using the USNOB1.0\footnote{{\tt http://www.nofs.navy.mil/data/fchpix/}}
catalogue. We grouped our images into 18 bins in order to increase the signal-to-noise
ratio (SNR) and performed aperture photometry with the SExtractor package \citep{Bertin96}
for all the objects in the field. In order to minimise the systematics, we performed
differential photometry with respect to a selection of local isolated and non-saturated
standard stars.



The calibration of NOT images taken with the $R$ filter was performed with respect to the
converted magnitudes in the $R$-band of the selected set of stars used for the calibration of
LT and GROND images in the SDSS passbands. We transformed the $r'$ and $i'$ magnitudes of the
calibration stars \citep{Cool08} into $R$ and $I$ magnitude following the filter
transformations of \citet{Krisciunas98}.

Hereafter, the magnitudes shown are not corrected for Galactic extinction, whilst
fluxes as well as all the best-fit models are. When the models are plotted together
with magnitudes, the correction for Galactic extinction is removed from the models.


\begin{table*}
 \begin{center}
 \caption{Best-fit parameters of the 15--150~keV and of the 0.3--10~keV spectra
of the $\gamma$-ray prompt and X-ray afterglow measured with BAT and XRT, respectively.
The model is an absorbed power law ({\tt xspec} model: {\tt pegpwrlw} for BAT and
{\tt wabs} {\tt pow} for XRT data, respectively). Frozen values are reported among
square brackets. Also SED modelling results are reported, both derived on the broad
band from optical to X-rays and optical points alone. The extinction is modelled with
a SMC profile as parametrised by \citet{Pei92}.}
 \label{t:BAT_XRT_spec}
 \begin{tabular}{lcrrccccc}
 \hline
 \hline
 \noalign{\smallskip}
 Interval  & Energy band & Start time  & Stop time  & $\beta$  & $N_{\rm H,z}$         & Mean flux            & $A_{V,z}^{a}$ & $\chi^2$/dof\\
           & (keV)       & (s)         & (s)        &          & ($10^{21}$~cm$^{-2}$) & (erg~cm$^{-2}$~s$^{-1}$) &       &  \\
\noalign{\smallskip}
\hline
\noalign{\smallskip}
$T_{90}$      &  15--150 &   $0.0$  &   $67.0$ & $1.44\pm0.46$    &  --  & $(3.3\pm0.8)\times10^{-7}$ &  --  & $1.44/6$ \\
Total         &  15--150 &  $-2.0$  &   $90.0$ & $1.65\pm0.51$    &  --  & $(3.6\pm0.8)\times10^{-7}$ &  --  & $5.6/7$  \\
Pulses 1--3   &  15--150 &  $-2.0$  &   $20.0$ & $2.0\pm0.5$      &  --  & $(2.2\pm0.5)\times10^{-7}$ &  --  & $1.82/7$ \\
Pulse 4       &  15--150 &  $52.9$  &   $90.0$ & $1.4\pm0.5$      &  --  & $(1.4\pm0.5)\times10^{-7}$ &  --  & $5.8/5$  \\
Peak          &  15--150 &  $0.384$ &   $0.832$& $1.1\pm0.6$      &  --  & $(1.0\pm0.2)^{b}$          &  --  & $6.1/6$  \\
\hline
XRT-WT  &  0.3--10 &  $77$  & $134$  & $1.06_{-0.09}^{+0.10}$&$<1.4$& $(4.1\pm0.3)\times10^{-10}$  & -- & $28.3/31$ \\
Plateau &  0.3--10 &  $423$ & $1507$ & $0.80_{-0.15}^{+0.16}$ & $1.6_{-1.5}^{+1.8}$ & $(2.3\pm0.3)\times10^{-11}$ & -- & $17.2/24$ \\
\hline
SED~2      &  opt--X  & $186.8$ & $269.4$  & $0.74\pm0.03$& $[2.7]$ & -- & $< 0.04$ & $8.2/6$ \\
SED~3      &  opt--X  & $423$   & $1507$   & $0.79\pm0.01$ & $2.7\pm0.8$ & -- & $< 0.02$ & $32/34$ \\
SED~4      &  opt     & $78117$ & $93620$  & $0.85\pm0.30$ & -- & -- & $0.10_{-0.06}^{+0.14}$ & $3.4/4$ \\
\hline
SED~2      &  opt     & $186.8$ & $269.4$  & $0.61\pm0.13$ & --  & -- & $[0]$ & $3.1/3$ \\
SED~3      &  opt     & $423$   & $1507$   & $0.74\pm0.05$ & --  & -- & $[0]$ & $7.9/8$ \\
SED~4      &  opt     & $78117$ & $93620$  & $1.05\pm0.06$ & --  & -- & $[0]$ & $4.8/5$ \\
\noalign{\smallskip}
 \hline
 \end{tabular}
 \end{center} 
\flushleft
$^{a}$ Rest-frame extinction obtained by modelling the SED with an SMC profile as parametrised by \citet{Pei92}.\\
$^{b}$ Peak photon flux in units of ph~cm$^{-2}$~s$^{-1}$.\\
\end{table*}

\subsection{Spectroscopy}
\label{sec:NOT_spec}
\begin{figure*}
\centering
\includegraphics[width=17cm]{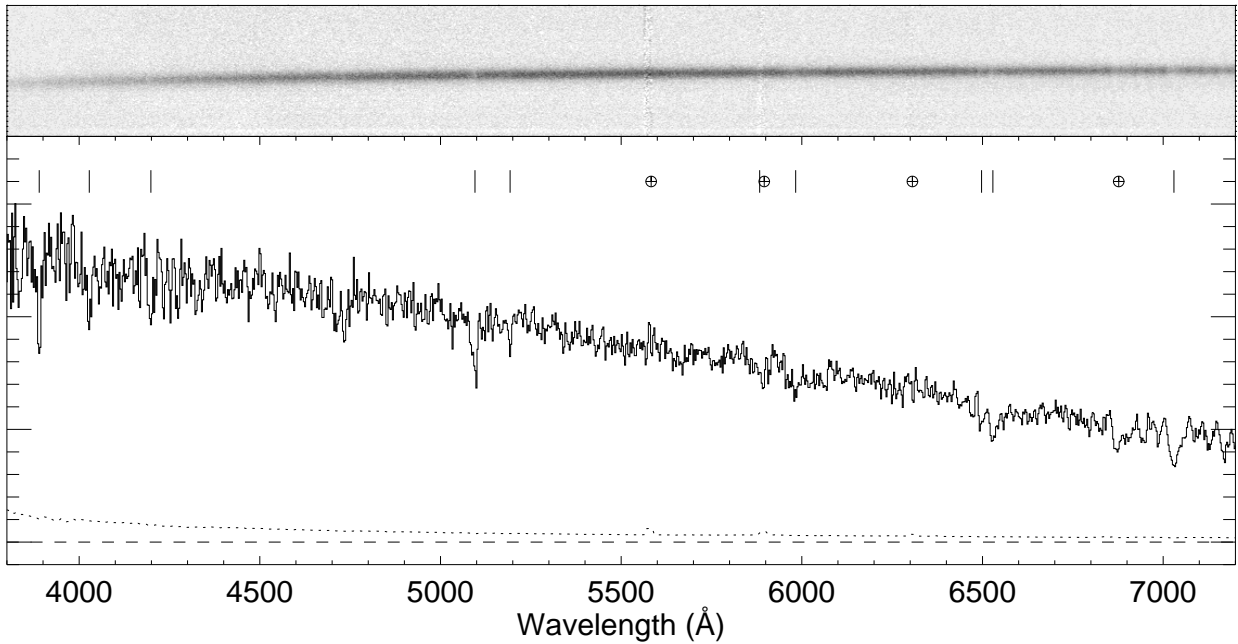}
\caption{The NOT spectrum of the afterglow taken at 46~min post burst. The tick marks and
Earth symbols show the absorption features (Table~\ref{t:spectrum}) and telluric lines, respectively.
The dotted line shows the error spectrum.}
\label{f:NOT_spec}
\end{figure*}

%
Starting at $\approx$$46$~min we obtained a 1800~s spectrum with a low resolution grism and a
1.3 arcsec wide slit covering the spectral range from about 3500 to 9000~{\AA} at a resolution of
14~{\AA} with the NOT (Fig.~\ref{f:NOT_spec}).
The airmass was about 1.8 at the start of the observations.  The spectrum was reduced using
standard methods for bias subtraction,  flat-fielding and wavelength calibration using an Helium-Neon
arc spectrum. The rms of the residuals in the wavelength calibration were about 0.3~{\AA}.
The spectrum was flux-calibrated using an observation  with the same setup of the spectrophotometric
standard star HD93521. Table~\ref{t:spectrum} reports the identified lines.
%
%

\section{Multi-wavelength combined analysis}
\label{sec:multi}

\subsection{Panchromatic light curve}
\label{sec:multi_lc}
\begin{figure*}
\centering
\includegraphics[]{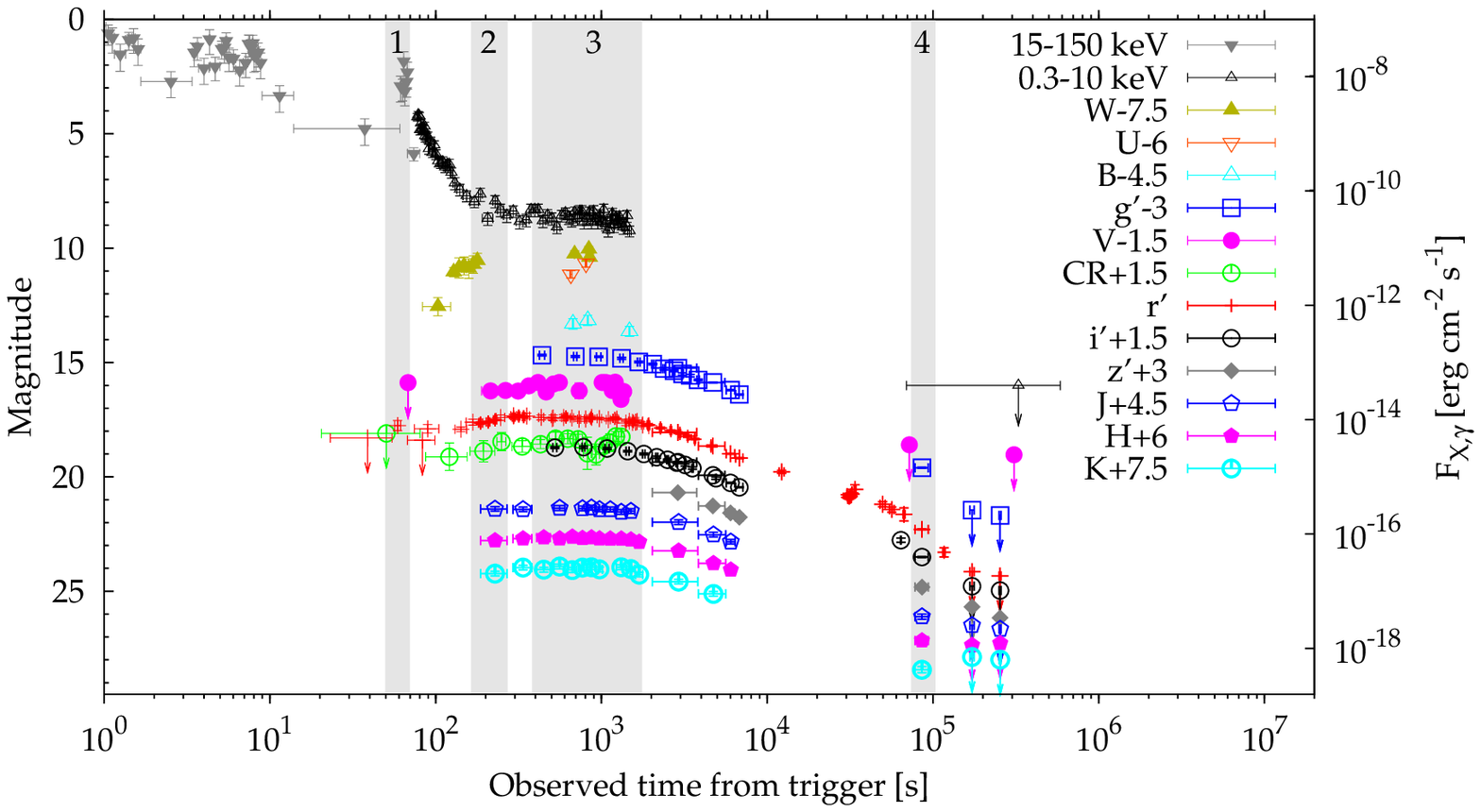}
\caption{Panchromatic light curve of the prompt (BAT, 15--150~keV), X-ray (XRT, 0.3--10~keV)
and NIR/visible/UV (FTN, GROND, LT, NOT, REM, TAROT, UVOT--W) afterglow of XRF~080330. Magnitudes (high-energy fluxes)
are referred to the left-hand (right-hand) y-axis. RAPTOR points were taken from \citet{Wren08} and
shifted by $0.3$~mag in order to match contemporaneous $r'$ points. The shaded bands indicate the four
intervals where we computed the SEDs.}
\label{f:multi_lc}
\end{figure*}

Figure~\ref{f:multi_lc} displays the light curves of the prompt emission (15--150~keV)
and of the 0.3--10~keV and NIR/visible/UV afterglow derived from our data sets plus some points
taken from RAPTOR \citep{Wren08}. High-energy fluxes (magnitudes) are referred to the
right-hand (left-hand) y-axis.
First of all, we note that the peak time of the last $\gamma$-ray pulse (Table~\ref{t:BAT_N05}) is
contemporaneous with the optical flash detected by RAPTOR, reported at $58.9\pm2.5$~s \citep{Wren08}.
%
%

The initial steep decay observed by XRT is a smooth continuation of the last $\gamma$-ray
pulse and is thus the tail of the prompt GRB emission, and likely to correspond
to its high-latitude emission. Most notably, during
the X-ray steep decay the optical flux is seen to rise up to $\sim$$300$~s and finally
a simultaneous plateau is reached at both energy bands,
lasting up to $\sim$$1500$~s, when the X-ray observations stopped.
This strongly suggests that the plateau is emission from a region which is physically
distinct from that responsible for the prompt emission and its tail (the rapid decay phase). 

As discussed in Sect.~\ref{sec:SED}, the afterglow does not show evidence for
spectral evolution throughout the observations, except for late epochs ($t\sim10^5$~s),
when there is evidence for reddening.
The achromatic nature of the afterglow light curve allows for a multi-wavelength
simultaneous fit of twelve light curves, where only the normalisations are left free to
vary independently from each other. We consider all of the available passbands: $K$, $H$, $J$,
$z'$, $i'$, $r'$, $V$, $g'$, $B$, $U$, $UWV1$ and X-ray, respectively. The latter
curve is fitted from 300~s onward, so as to exclude the initial steep decay.
Hereafter we present two alternative combinations of models, both providing a reasonable
description of the flux temporal evolution.
In both cases we had to add a 2\% systematics to all of the measured uncertainties
to account for some residual variability with respect to the models, in order to have
acceptable $\chi^2$ values and correspondingly acceptable parameters' uncertainties.

\subsubsection{Multiple smoothly broken power law}
\label{sec:beu3}
\begin{figure*}
\centering
\includegraphics{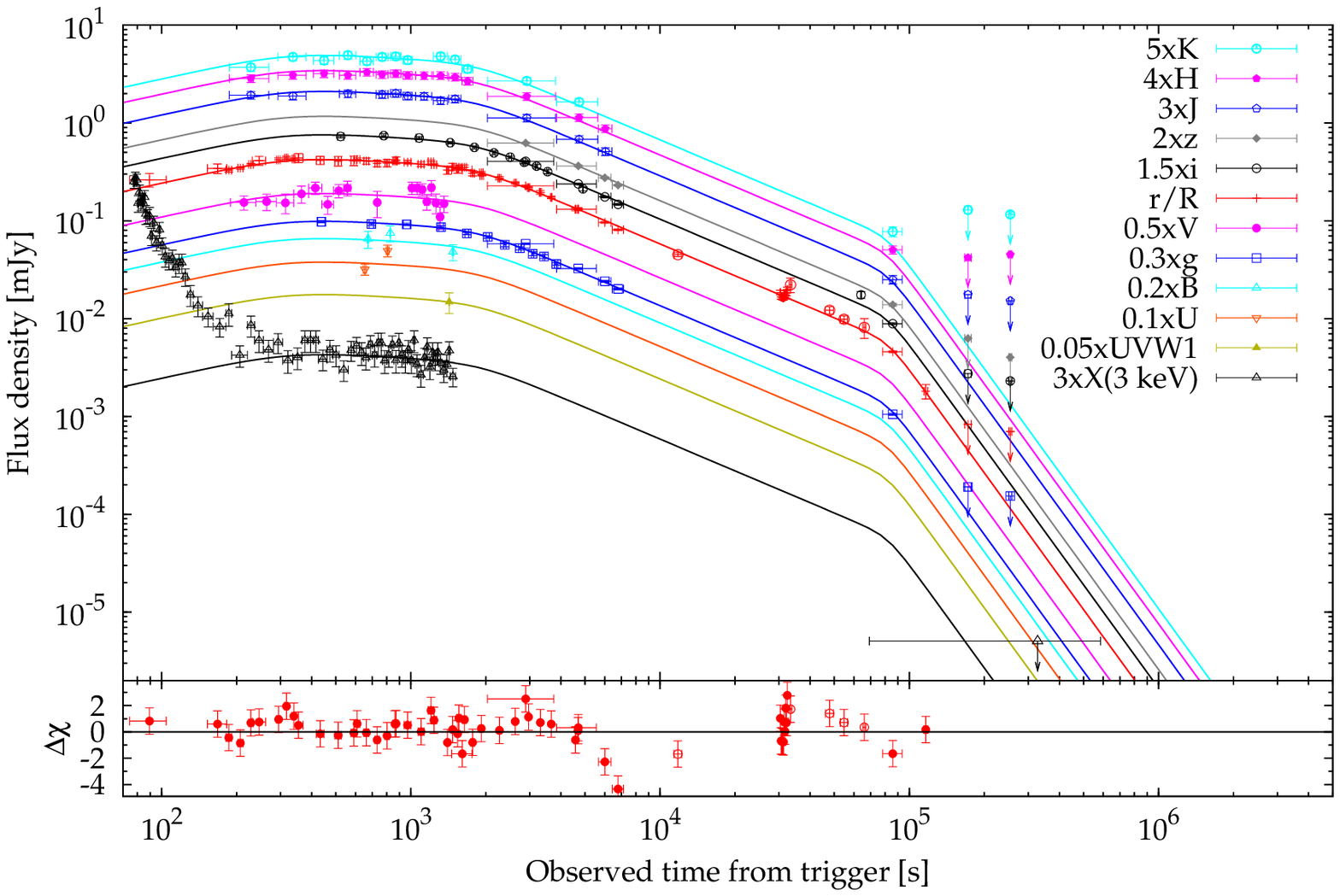}
\caption{{\em Top panel}: NIR/visible/UV/X-ray light curves expressed in flux densities units, after
correction for Galactic extinction. The effective wavelength decreases from top to bottom, from
$K$ filter all the way down to X-rays. Red empty circles are public data points from literature
\citep{Im08,Wang08,Sergeev08,Moskvitin08}.
Upper limits are 3$\sigma$. The solid lines show the same model of a multiple smoothly
broken power-law obtained by fitting all the curves simultaneously, allowing for different
normalisations but with the same fit parameters. {\em Bottom panel}: residuals of the $r'$ curve
with respect to the corresponding model.}
\label{f:fluxall_beu3}
\end{figure*}

A possible description of the light curves is offered by a multiple broken power law
(Fig.~\ref{f:fluxall_beu3}).
This has the advantage of a more straightforward interpretation in terms of the standard
fireball evolution model due to synchrotron emission. We started from the parametrisation
by \citet{Beuermann99} and added two more breaks to finally provide a sufficiently detailed
description. The fitting function is given by eq.~(\ref{eq:beu3}).
\begin{equation}
\displaystyle F(t)\,= \,\frac{F_0}{\left[\left(t/t_{b1}\right)^{n\,\alpha_1} +
\left(t/t_{b1}\right)^{n\,\alpha_2} + \left(t/t_{b2}\right)^{n\,\alpha_3}
+ \left(t/t_{b3}\right)^{n\,\alpha_4} \right]^{1/n}}
\label{eq:beu3}
\end{equation}
The free parameters are the normalisation constant (different for each curve), $F_0$,
three break time constants,  $t_{bi}$ ($i=1,2,3$), four power-law indices,
$\alpha_1<\alpha_2<\alpha_3<\alpha_4$, the smoothness $n$. Apart from the normalisations,
all the curves share the same parameters. Overall, the free parameters and the degrees of
freedom (dof) total 20 and 184, respectively.
%
%
Equation~(\ref{eq:beu3}) looks like a piecewise power law only in the following
regime: $t_{b1}\ll\,t_{b2}\ll\,t_{b3}$, where each individual term takes over at well separate
epochs. The light curve of XRF~080330 fits in this case, as proven by the best-fit results
(first line of Table~\ref{t:panchro_fit}) and shown in Fig.~\ref{f:fluxall_beu3}.
The effective break times, $t_{b1,{\rm eff}}$, $t_{b2,{\rm eff}}$, $t_{b3,{\rm eff}}$, i.e.
the times at which the model~(\ref{eq:beu3}) changes the power-law regime, are simply
given by $t_{b1,{\rm eff}}=t_{b1}$,
$t_{b2,{\rm eff}}=(t_{b1}^{\alpha_2}/t_{b2}^{\alpha_3})^{1/(\alpha_2-\alpha_3)}$,
$t_{b3,{\rm eff}}=(t_{b2}^{\alpha_3}/t_{b3}^{\alpha_4})^{1/(\alpha_3-\alpha_4)}$.

The goodness of the fit in terms of $\chi^2/{\rm dof}$ is 212/184, corresponding to a
non-rejectable P-value of $7.7$\%. The normalisation constants for the different bands
are the following ($\mu$Jy): $F_K=1077_{-59}^{+54}$, $F_H=942_{-50}^{+45}$, $F_J=769_{-44}^{+40}$,
$F_z=643_{-30}^{+26}$, $F_i=554_{-25}^{+19}$, $F_r=464_{-21}^{+16}$, $F_V=418_{-33}^{+34}$,
$F_g=362_{-16}^{+12}$, $F_B=362\pm68$, $F_U=277\pm45$, $F_{UVW1}=129\pm52$, while the
X-ray normalisation, expressed in flux units in the 0.3--10~keV band instead of flux density, is
$F_{\rm x}=(3.7\pm0.3)\times10^{-11}$~erg~cm$^{-2}$~s$^{-1}$.
The effective break times are found be $t_{b1,{\rm eff}}=317$~s, $t_{b2,{\rm eff}}=1850$~s
and $t_{b3,{\rm eff}}=82.4$~ks, respectively.

The bottom panel of Fig.~\ref{f:fluxall_beu3} shows the residuals of the $r'$ curve with
respect to the model; the displayed uncertainties do not include the 2\% systematics added
by the fitting procedure. We note that between $6$ and $7\times10^3$~s the model overpredicts
the flux by 2--4$\sigma$ with respect to the measured values, corresponding to a $\sim\,0.1$
magnitude difference. However, the later points seem to rule out a steeper decay than the modelled one.
Alternatively, one might interpret this as suggestive of a steeper decay followed by a second
component thus causing a late flux enhancement.
This possibility motivated us to provide an alternative description, described in the next
Sect.~\ref{sec:beulore}.

\subsubsection{A two-component model: late-time brightening}
\label{sec:beulore}
\begin{figure*}
\centering
\includegraphics{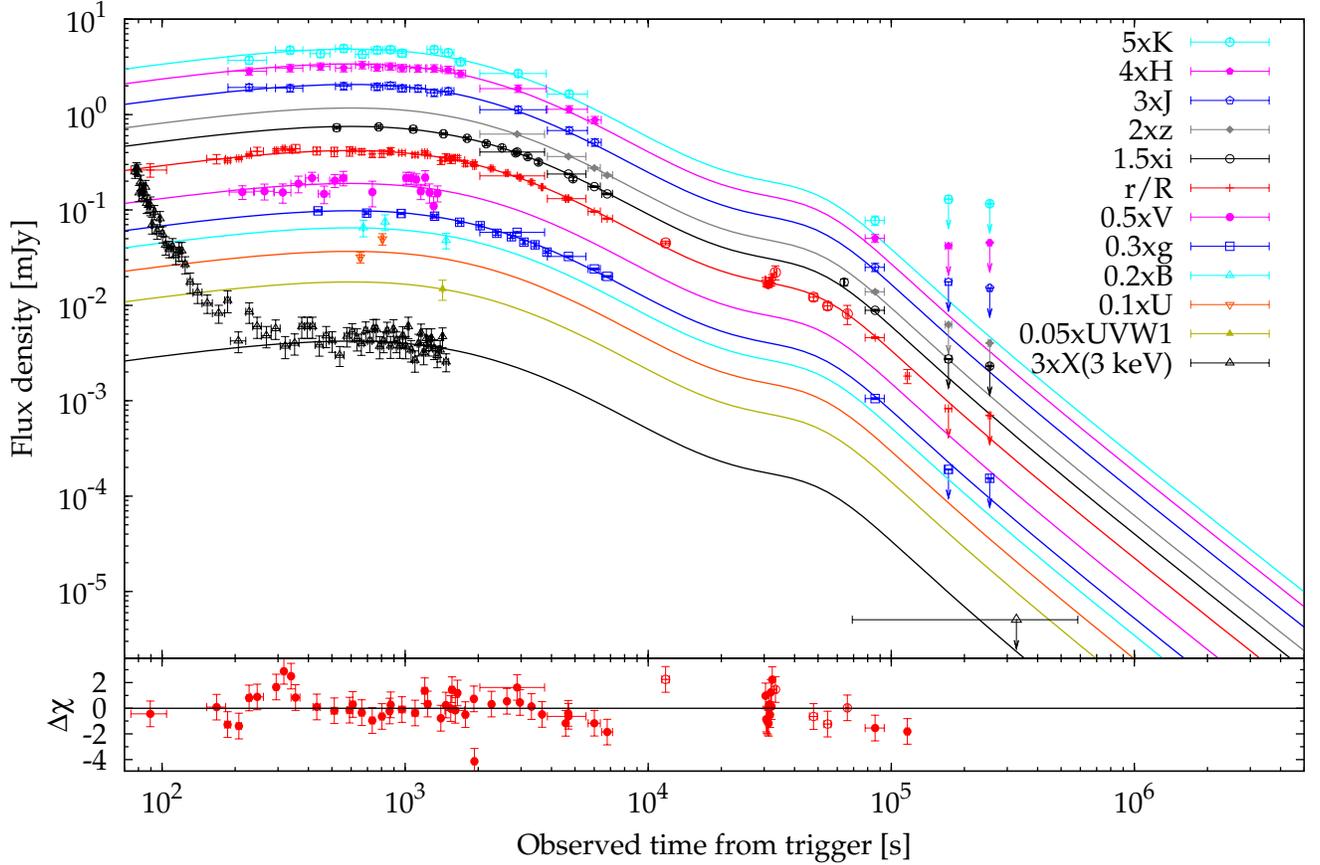}
\caption{{\em Top panel}: same as Fig.~\ref{f:fluxall_beu3}. In this case the solid lines show the same model
of a smoothly broken power-law plus a Lorentzian obtained by fitting all the curves simultaneously,
allowing for different normalisations but with the same fit parameters.
{\em Bottom panel}: residuals of the $r'$ curve with respect to the corresponding model.}
\label{f:fluxall_beulore}
\end{figure*}

In Figure~\ref{f:fluxall_beulore} we modelled the first part ($t<10^4$~s)
with a simple smoothly broken power law with a single break time, $t_{b1}$,
and two power-law indices, $\alpha_1$ ($\alpha_2$), taking over for
at $t\ll t_{b1}$ ($t\gg t_{b1}$). In order to model the later data points, we had to add
a further component. A Lorentzian proved successful in this respect, so that the complete
model used is given by the following equation.
\begin{equation}
\displaystyle F(t)\,= \,\frac{F_{0,r}}{\left[\left(t/t_{b1}\right)^{n\,\alpha_1} +
\left(t/t_{b1}\right)^{n\,\alpha_2}\right]^{1/n}}\, + \,
\frac{F_{\rm {\sc L},r}}{1 + \left[2\left(t-t_{\rm c}\right)/t_{\rm w}\right]^2}
\label{eq:beulore}
\end{equation}
This was used to fit the $r'$ curve. 
The free parameters are the normalisation constant, $F_{0,r}$,
the break time, $t_{b1}$, two power-law indices, $\alpha_1$ and $\alpha_2$,
the smoothness $n$, the Lorentzian normalisation, $F_{\rm {\sc L},r}$,
the peak time, $t_{\rm c}$ and its width, $t_{\rm w}$. The time-integrated
flux density of the latter component is $\pi\,F_{\rm {\sc L},r}\,t_{\rm w}/2$.
The two terms of eq.~(\ref{eq:beulore}) peak at
$t_{p1}=t_{b1}(-\alpha_1/\alpha_2)^{1/[n(\alpha_2-\alpha_1)]}$ and
$t_{p2}=t_{\rm c}$, respectively.
Each of the light curves of the remaining filters were fitted with a free scaling
factor with respect to the $r'$ curve as modelled by eq.~(\ref{eq:beulore}). 
The free parameters and the dof total 19 and 185, respectively.
%
%
The best-fit result is shown in Fig.~\ref{f:fluxall_beulore}, while the second line of
Table~\ref{t:panchro_fit} reports the corresponding best fit values.
The fit is good: $\chi^2/{\rm dof}=187/185$. The scaling factors for the remaining bands
are the following: $f_K=2.31\pm0.12$, $f_H=2.03\pm0.10$, $f_J=1.65\pm0.09$,
$f_z=1.40\pm0.04$, $f_i=1.20\pm0.03$, $f_V=0.90\pm0.08$, $f_g=0.78\pm0.02$,
$f_B=0.77_{-0.15}^{+0.19}$, $f_U=0.59_{-0.10}^{+0.12}$, $f_{UVW1}=0.28_{-0.11}^{+0.17}$,
while the X-ray normalisation is still
$F_{\rm x}=(3.7\pm0.3)\times10^{-11}$~erg~cm$^{-2}$~s$^{-1}$.
The two components peak at $t_{p1}=600$~s and $t_{p2}=34.4$~ks, respectively.

We also tried to model the second component with a rising and falling smoothly broken power law instead of
a Lorentzian. However, this brings in too many free parameters, such as the slope of the rise, so unless
one finds reasons to fix some of them to precise values, the fit with such a component turns into highly
undetermined parameters.

\begin{table*}
 \begin{center}
 \caption{Best-fit parameters of the multi-wavelength fitting procedure of the afterglow light curves.}
 \label{t:panchro_fit}
 \begin{tabular}{rrrrrrrrrrrr}
 \hline
 \hline
 \noalign{\smallskip}
$\alpha_1$ & $t_{b1}$ & $\alpha_2$  & $t_{b2}$  & $\alpha_3$  & $t_{b3}$ & $\alpha_4$ & $n$ & $t_{\rm c}$
                 & $t_{\rm w}$ & $F_{\rm {\sc L},r}$ & $\chi^2$/dof\\
                 & (s)      &             &   (s)     &             & (ks)     &            &     & (ks)
                 &   (ks)      &     ($\mu$Jy)         &             \\
\hline
$-0.56_{-0.33}^{+0.24}$ & $317_{-76}^{+151}$ & $0.15_{-0.07}^{+0.09}$ & $1456_{-46}^{+67}$ &
                 $1.08\pm0.02$ & $23.8_{-3.1}^{+3.2}$ & $3.51_{-0.34}^{+0.37}$ & $5.4_{-1.3}^{+1.9}$
                 & -- & -- & -- & $212/184$\\
$-0.38_{-0.23}^{+0.22}$ & $2480_{-900}^{+1420}$ & $2.02_{-0.75}^{+0.85}$ & -- &
                 -- & -- & -- & $0.49_{-0.28}^{+0.61}$
                 & $34.4_{-8.1}^{+10.6}$ & $72.7_{-12.2}^{+14.6}$ & $11.9_{-2.7}^{+3.5}$ & $187/185$\\
\noalign{\smallskip}
 \hline
 \end{tabular}
 \end{center} 
\end{table*}

\subsection{Spectral Energy Distribution}
\label{sec:SED}
\begin{figure}
\centering
\includegraphics[width=8.5cm]{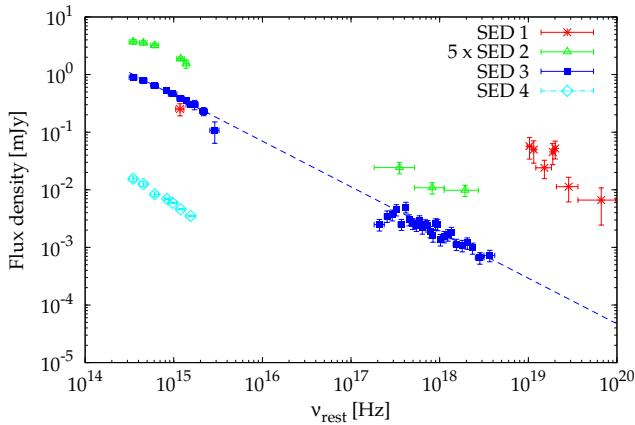}
\caption{GRB rest-frame SEDs 1 to 4 (shown with asterisks, triangles, squares and diamonds, respectively).
The dashed line shows the best-fitting power-law model of SED~2: $\beta_{\rm ox}=0.79\pm0.01$ and
$A_{V,z}<0.02$. X-ray data are not absorption-corrected.}
\label{f:SEDall}
\end{figure}

Figure~\ref{f:SEDall} displays four SEDs we derived in as many different time intervals
(see shaded bands in Fig.~\ref{f:multi_lc}):
\begin{enumerate}
\item SED~1 includes the last $\gamma$-ray pulse and the optical flash detected by RAPTOR
      \citep{Wren08}, around 60~s;
\item SED~2 corresponds to the final part of the optical rise, coinciding with the final
      part of the X-ray steep decay, spanning from $186.8$ to $269.4$~s;
\item SED~3 has the broadest wavelength coverage and corresponds to the plateau phase,
      from $\sim$$400$ to $\sim$$1500$~s.
\item SED~4 includes NIR/visible measurements around the possible late time break in the light
      curve (Fig.~\ref{f:fluxall_beu3}), at~$\sim$$10^5$~s.
\end{enumerate}

%
%

\begin{figure}
\centering
\includegraphics[height=8.5cm,angle=270]{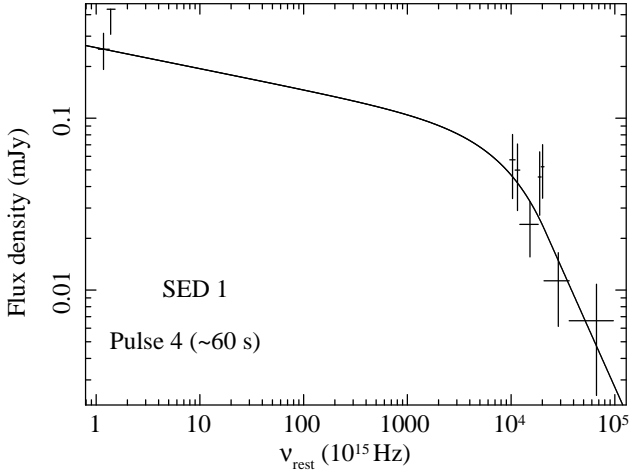}
\caption{GRB rest-frame SED~1 from observed optical to $\gamma$-ray during the optical flash concomitant
with the last $\gamma$-ray pulse at $\sim60$~s.
The solid line shows the best-fitting smoothed broken power law \citep{Band93} with the following
parameters: $\alpha_{\rm B}=-1.1$, $\beta_{\rm B}=-2.35$ and $E_{\rm p,i}=71$~keV.}
\label{f:SED1}
\end{figure}

To construct SED~1 we made use of the RAPTOR measurement \citep{Wren08}, a UVOT upper limit
in the $V$ band and of the BAT spectrum of the fourth pulse.
Figure~\ref{f:SED1} displays this SED: the solid line shows the best fit with a smoothed
broken power law  used to fit the high-energy photon spectra of the prompt emission of GRBs
\citep{Band93}.
%
%
The best-fitting parameters are the following: $\alpha_{\rm B}=-1.12$, $\beta_{\rm B}=-2.35$
and $E_{\rm p,i}=71$~keV ($\chi^2/{\rm dof}=5.8/5$) consistent with
the limit on $E_{\rm p,i}$ derived in Sect.~\ref{sec:gamma}. The Band indices are photon indices, so
the corresponding energy indices are $0.12$ and $1.35$, respectively.
While $\beta_{\rm B}$ was constrained by the BAT data themselves, we solved the coupled indetermination
$\alpha_{\rm B}$--$E_{\rm p,i}$
by initially freezing the low-energy index $\alpha_{\rm B}$ to the typical value of
$-1$ \citep{Kaneko06} and then leaving it to vary. The above minimum $\chi^2$ was so found.
Although this does not break the degeneracy of both parameters (for every
$E_{\rm p,i}<88$~keV there is a value of $\alpha_{\rm B}$ for which an acceptable fit is given), here it
is shown that the extrapolation of a typical Band model fitting the spectrum of the last pulse matches
the optical flux observed during the flash. However, because of the lack of measurement of $\alpha_{\rm B}$
from $\gamma$-ray data, the optical flux matched by the extrapolation of the Band model may still be accidental.

The time interval of SED~2 corresponds to the first $JHK$ GROND frames
and spans from $186.8$ to $269.4$~s (see Fig.~\ref{f:multi_lc}).
It consists of contemporaneous $Vr'JHK$ frames as well as X-rays.
%
\begin{figure}
\centering
\includegraphics[height=8.5cm,angle=270]{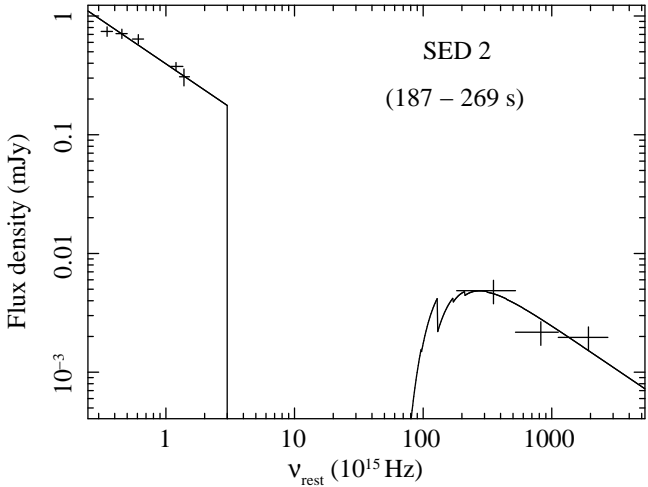}
\caption{GRB rest-frame SED~2 corresponding to the final part of the rise of the optical afterglow.}
\label{f:SED2}
\end{figure}
%
The NIR-to-X SED~2 can be fitted with a single power-law with $\beta_{\rm ox}=0.74\pm0.03$ and
negligible dust extinction. The optical data alone can be fitted with an unextinguished power
law with $\beta_{\rm o}=0.61\pm0.13$ (Table~\ref{t:BAT_XRT_spec}).

SED~3, taken during the plateau, is the richest one including all of the passbands considered
in this work, but the $\gamma$-rays (see Fig.~\ref{f:multi_lc}).
Our NIR values are consistent with the $JHK$ points of
\citet{Bloom08}. Given the steadiness of the light curve and the evidence for no
significant colour change along the plateau, the SED so obtained is fairly robust.
%
\begin{figure}
\centering
\includegraphics[height=8.5cm,angle=270]{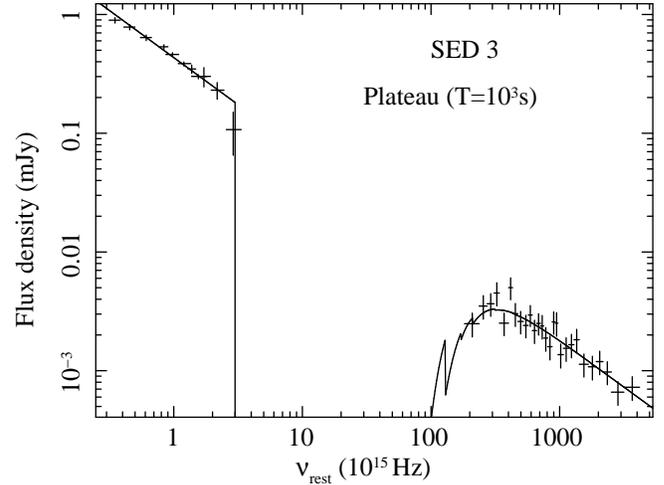}
\caption{GRB rest-frame SED from observed NIR to X-ray during the plateau from $400$ to $1500$~s.
The solid line shows the best-fit model (SMC profile) corresponding to a single unextinguished
power law with spectral index $\beta=0.79\pm0.01$ and negligible local-frame extinction,
$A_{V,z}<0.02$ (see text).}
\label{f:SED3}
\end{figure}
%
The multi-wavelength fitting of the light curves of Figs.~\ref{f:fluxall_beu3} and
\ref{f:fluxall_beulore} (Sect.~\ref{sec:multi_lc}) is dominated by the data points along the plateau phase.
Thus, we built a SED using the best-fit normalisations
(Sects.~\ref{sec:beu3} and \ref{sec:beulore}) and calculated at the time of $10^3$~s.
We found the same results with improved uncertainties, due to the stronger constraints
imposed by a multi-band fitting. The result of this SED is displayed in Figure~\ref{f:SED3}.
The solid line shows the best-fit model obtained adopting a rest-frame SMC-extinguished
\citep{Pei92}, X-ray photoelectrically absorbed power law with $\beta_{\rm ox}=0.79\pm0.01$.
We note that the point corresponding to the UVW1 filter nicely agrees with the Lyman
absorption at the GRB redshift.
The rest-frame optical extinction was found to be negligible and a very
tight limit could be derived, $A_{V,z}<0.02$.
Thanks to the more precise estimate obtained on $\beta_{\rm ox}$, the estimate 
of $N_{{\rm H},z}$ improved correspondingly: $(2.7\pm0.8)\times10^{21}$~cm$^{-2}$.
These results are consistent with what was obtained from the 0.3--10~keV spectrum alone and
the accuracy of the estimates benefited significantly from the inclusion of NIR/visible data.
Fitting the optical data alone, the result is similar: no need for a significant amount
of extinction and, more importantly, the same index: $\beta_{\rm o}=\beta_{\rm ox}=\beta_{\rm x}$
(Table~\ref{t:BAT_XRT_spec}), thus ruling out the possibility of a significant reddening
due to dust along the line of sight within the host galaxy.

As a consequence, two properties are inferred:
\begin{itemize}
\item a negligible dust column density in the circumburst environment and along the line of sight
to the GRB through the host galaxy;
\item a single power-law component accounting for the (observed) NIR to X-ray radiation, pointing
to a single emission mechanism with no breaks in between. Furthermore, a single power-law spectrum
implies an achromatic evolution, consistently with the observations, while the other way around
is not true.
\end{itemize}

The epoch of SED~4 is $\sim$$80$~ks, i.e. around the final break (aftermath of the late brightening) 
following the light curve description given in Sect.~\ref{sec:beu3} (Sect.~\ref{sec:beulore}) and
shown in Fig.~\ref{f:fluxall_beu3} (Fig.~\ref{f:fluxall_beulore}).
This includes optical data and is shown in Fig.~\ref{f:SEDall} (diamonds).
Data can be fitted either with a single unextinguished power-law with $\beta_{\rm o}=1.05\pm0.06$ ($A_{V,z}$
fixed to 0) or, alternatively, with $\beta_{\rm o}=0.85\pm0.30$ and some extinction,
$A_{V,z}=0.10_{-0.06}^{+0.14}$. An increase of the extinction along with time seems hard to explain
physically, so we are led to favour a true reddening at this time.
Compared with the previous SEDs (Table~\ref{t:BAT_XRT_spec}), SED~4 is redder by
$\Delta\beta_{\rm o}=0.26\pm0.06$ (significance of $\sim$$6\times10^{-5}$).
We point out that the reddening is independent of the fit choice, as demonstrated from the comparison
of the bare power-law indices with no dust correction between the earlier and later spectra
(Table~\ref{t:BAT_XRT_spec}).

\section{Discussion}
\label{sec:disc}
The 15--150~keV fluence and peak flux of XRF~080330 are typical of other XRFs detected by Swift.
The X-ray afterglow flux places XRF~080330 in the low end of the distribution of
the GRBs observed by Swift, similarly to the majority of XRFs \citep{Sakamoto08}.
Moreover, the observed X-ray flux of XRF~080330 lies in the low end of
both the XRFs sample of Swift considered by \citet{Sakamoto08}  and of the XRFs sample
of BeppoSAX of \citet{Dalessio06}. 
The X-ray afterglow of XRF~080330 does require a remarkable steepening after the shallow phase,
with $\alpha_{\rm x,3}>1.3$ regardless of the light curve modelling (Fig.~\ref{f:multi_lc}).
This decay is typical of classical GRBs (or steeper), but is in contrast to the fairly shallow
decays found by \citet{Sakamoto08} for their sample of Swift XRFs.
The optical flux of the XRF~080330 afterglow is within 1$\sigma$ of the distribution of the
BeppoSAX XRFs sample of \citet{Dalessio06} at 40~ks post burst.

The coincidence of the steep decay observed in the X-ray light curve,
that is a smooth continuation of the last $\gamma$-ray pulse,
suggests that this corresponds to its high-latitude emission or the so-called
``curvature effect'' \citep{Fenimore96,Kumar00,Dermer04}. In the case of a thin shell emitting
for a short time, the closure relation expected between temporal and spectral indices is 
$\alpha=\beta+2$, with the time origin $t_0$ reset to the ejection time of the related pulse,
earlier than the onset by about 3--4 times the width of the pulse.
This still holds even if the emission occurs over a finite range of radii \citep{Genet08},
though in that case the ratio of the ejection to pulse onset time difference and the pulse
width becomes smaller ($\sim\,1$ for $\Delta\,R\sim\,R$).

\begin{figure}
\centering
\includegraphics[width=8.5cm]{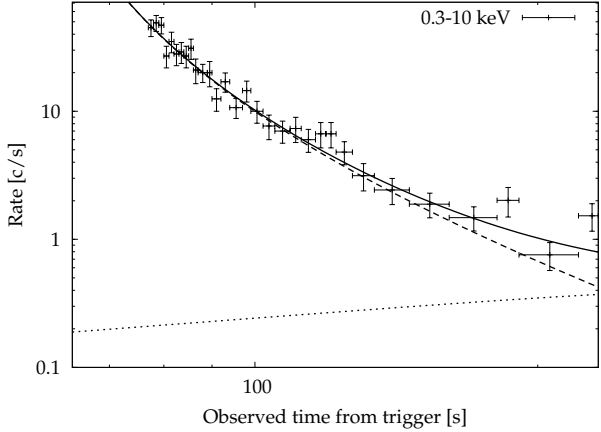}
\caption{0.3--10~keV steep decay curve. The solid line shows the best-fitting
model of the form $F(t)+k\,(t-t_0)^{-\alpha_{\rm x}}$, where the parameters of $F(t)$ (eq.~\ref{eq:beu3})
were frozen to their best-fit values obtained by the multi-band fit of the afterglow component
(Sect.~\ref{sec:beu3}) and with $k$, $t_0$ and $\alpha_{\rm x}$
free parameters. Both terms, $F(t)$ and the power-law, are shown separately
by the dotted and dashed curves, respectively.}
\label{f:xrt_curvature}
\end{figure}

We fitted the X-ray decay up to $210$~s with the form $F(t)+{\bf k\,}(t-t_0)^{-\alpha_{\rm x}}$, where
the parameters of $F(t)$ (eq.~\ref{eq:beu3}) were frozen to their corresponding best-fit values
obtained in Sect.~\ref{sec:beu3}, and the power-law parameters $t_0$, $\alpha_{\rm x}$ and its
normalisation ${\bf k}$ were left free to vary. We obtained: $t_0=53_{-18}^{+9}$~s and
$\alpha_{\rm x}=2.4_{-0.5}^{+0.9}$~s ($\chi^2/{\rm dof}=22/27$), as shown by the solid
line in Fig.~\ref{f:xrt_curvature}. During the steep decay, it is
$\beta_{\rm x}=1.06_{-0.09}^{+0.10}$ (Table~\ref{t:BAT_XRT_spec}). The curvature
relation is fully satisfied and we note that $t_0$ does correspond to the time
of the last pulse.
Replacing eq.~(\ref{eq:beu3}) with eq.~(\ref{eq:beulore}) for the underlying component, $F(t)$,
the best-fit parameters do not change to a noticeable degree. 
%
%

The optical afterglow of XRF~080330 exhibited a slow rise up to $\sim\,300$~s, followed by
plateau out to $\sim\,2\times\,10^3$~s, after which it decayed within a typical power-law index
of about $1.1$ approximately out to a few $10^4$~s. Then, a sharp break to a decay
index of $3.5_{-0.3}^{+0.4}$ occurred concurrently with an optical reddening
(Sect.~\ref{sec:beu3}; Fig.~\ref{f:fluxall_beu3}). Alternatively, after the plateau
a more gradual transition to a power-law decay index of $2.0\pm0.8$ set in, followed by a smooth,
red bump (Sect.~\ref{sec:beulore}; Fig.~\ref{f:fluxall_beulore}).
We discuss each phase separately in the following subsections.

\subsection{Optical afterglow rise}
\label{sec:rise}
In the context of the fireball model (e.g. M\'esz\'aros 2006 and references therein),
\nocite{Meszaros06} the possibility that the peak of the optical afterglow emission
corresponds to the passage of the peak synchrotron frequency is ruled out by the lack
of spectral evolution:
$\beta_{\rm o}$ should evolve from negative ($-1/3$) to positive values, while we find
no evidence for $\beta_{\rm o}$ changing before $\sim$$8\times10^4$~s.

Another possible interpretation of the optical peak is the onset of the afterglow, as
for GRB~060418 and GRB~060607A \citep{Molinari07,Jin07} and, possibly, for XRF~071010A
as well \citep{Covino08}. In the case of XRF~080330, the rise during the pre-deceleration
of the fireball within an ISM is much shallower than $\alpha\sim-3$, expected at frequencies
between $\nu_{\rm m}$ and $\nu_{\rm c}$ \citep{Sari99,Granot05,Jin07}. A wind environment would fit
in a better way the slow rise of XRF~080330. Under these assumptions and in the thin
shell case as the duration of the GRB is much shorter than the deceleration time,
we can estimate the initial Lorentz factor, $\Gamma_0$ (approximately, twice as large as the Lorentz
factor at the peak), in a wind-shaped density profile, $n(r)=A\,r^{-s}$ ($A$ is constant),
with $s=2$, from the peak time and the $\gamma$-ray
radiated energy, $E_{\rm iso}$ \citep{Chevalier00,Molinari07}. 
For consistency, only the two-component model (Sect.~\ref{sec:beulore};
Fig.~\ref{f:fluxall_beulore}) must be considered.
In this case, we take the peak time of the first component,
$t_{p1}=600$~s: assuming $\eta=0.2$ (radiative efficiency), $A=3\times10^{35}$~cm$^{-1}$,
it turns out $\Gamma_0<80$, and a corresponding deceleration radius smaller than
$7\times10^{16}$~cm. As in the case of XRF~071010A, the initial
Lorentz factor is smaller than those found for classical GRBs.
There is no evidence for the presence of a reverse shock; should the injection frequency of
the reverse shock lie within the optical passbands, it would dominate the optical flux
and exhibit a fast ($\sim$$t^{-2.1}$) decay \citep{Kobayashi07}, not observed here.
Nonetheless, this can still be the case if the injection frequency lies below
the optical bands \citep{Jin07,Mundell07}. 
A weak point of this interpretation is that the case $s=2$ is ruled out:
$\alpha=s(p+5)/4-3=0.79\pm0.01$ \citep{Granot05}. Inverting this relation, a value
of $s=1.4\pm0.1$ is required to explain the observed $\alpha=-0.4\pm0.2$.
This argument, together with the absence of reverse shock, whose $F_{\nu,{\rm max}}$ should
be much larger compared with the forward shock by a factor of $\sim\,\Gamma$ (although see
above), makes the interpretation of a deceleration through a wind environment somewhat contrived.

In the context of a single jet viewed off-axis \citep{Granot05_2}, the rising part of the
XRF~080330 curve is explained by the emission coming from the edge of the jet: as the bulk Lorentz
factor $\Gamma$ decreases, the beaming cone gets progressively wider, thus resulting in a rising flux.
The peak in the light curve is reached when it is $\Gamma\,\sim\,1/(\theta_{\rm obs}-\theta_0)$,
where $\theta_{\rm obs}$ and $\theta_0$ are the viewing and jet opening angles, respectively.
According to the optical afterglow classification given by \citet{Panaitescu08}, XRF~080330 belongs
to the class of slow-rising and peaking after 100~s events. Those authors found a possible anti-correlation
between the peak flux and the peak time for a number of fast-rising afterglows, followed
also by the slow-rising class and, in this respect, XRF~080330 is no exception.
The suggested interpretation of the rise is either the pre-deceleration synchrotron emission or
the emergence of a highly collimated outflow seen off-axis. In the latter case, assuming a power-law
angular distribution of the kinetic energy, $\mathcal{E}(\theta)\propto(\theta/\theta_{\rm c})^{-q}$ ($q>0$),
high values for $q$ correspond to slower rises and dimmer peak fluxes, for a fixed off-axis viewing
angle ($\theta_{\rm obs}=2\,\theta_{\rm c}$). In the former case, the anti-correlation is ascribed
to different circumburst environment densities for different events: XRF~080330,
because of the negligible dust extinction, would lie in the high-peak flux region, which is not
the case. This favours the interpretation of an off-axis jet whose angular distribution of energy
quickly drops away from the jet axis.

\begin{figure}
\centering
\includegraphics[width=8.5cm]{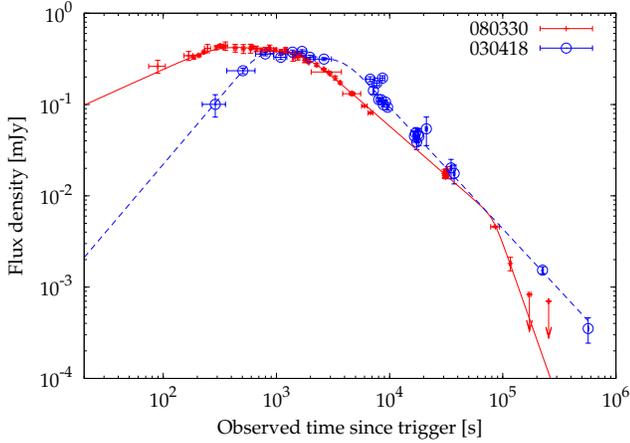}
\caption{$r$-band afterglow of XRF~080330 compared with the XRR~030418 (empty circles),
for which there is only an upper limit to its redshift, $z<5$ \citep{Dullighan03}.
Data of XRR~030418 have been taken from \citet{Rykoff04,Ferrero03,Dullighan03}.
The solid line is the best fit of the XRF~080330 $r$ curve of
Sect.~\ref{sec:beu3}, while the dashed line is the best fit obtained with the same
model applied to XRR~030418.}
\label{f:cfr_030418}
\end{figure}

An example of another XRF whose optical counterpart showed a very similar behaviour
is XRR~030418. The rise of this XRR, for which only an upper limit to its redshift ($z<5$)
was obtained \citep{Dullighan03}, has been interpreted as due to the decreasing extinction
along with time, caused by the dust column density crossed by the fireball
during its expansion \citep{Rykoff04}.
Figure~\ref{f:cfr_030418} shows the light curve compared with that of XRF~080330.
%
%
The solid line shows the best fit to the $r'$ curve of XRF~080330 of Sect.~\ref{sec:beu3}, while
the dashed line shows the same model applied to the XRR~030418 data.
XRR~030418 displays a steeper rise ($\alpha_1=-1.5$), which strongly depends on the zero
time and could be the same as that of XRF~080330 if the time origin was moved by $(190\pm50)$~s forward
in time (lab frame). However, there is nothing around this time in the $\gamma$-ray light curve of
XRR~030418.
Apart from the different slopes of the rise and the lack of a late-time steepening in the case of
XRR~030418, the plateau and post-plateau decay look very similar.
If both XRFs are caused by the same process, from the spectral (lack of) evolution XRF~080330 during
the rise-plateau-initial decay phases we can rule out the decreasing dust column density hypothesis.

\subsection{Plateau}
\label{sec:plateau}
From the SED extracted around the plateau no break is found between optical and X-ray
frequencies, with $\beta_{\rm ox}=0.79\pm0.01$. In the regime of slow cooling it is reasonable
to assume that both optical ($\nu_{\rm o}$) and X-ray (${\nu_{\rm x}}$) frequencies lie
between the injection ($\nu_{\rm m}$) and the cooling  ($\nu_{\rm c}$) frequencies:
$\nu_{\rm m}<\nu_{\rm o}<\nu_{\rm x}<\nu_{\rm c}$ \citep{Sari98}.
The power-law index of the electron energy distribution, $p$, is given by $\beta_{\rm ox}=(p-1)/2$,
yielding $p=2.58\pm0.02$, fully within the range of values found for other bursts
(e.g. Starling et~al. 2008).\nocite{Starling08}
The temporal decay index depends on the density profile: the ISM (wind) case predicts a
value of $\alpha=3(p-1)/4=1.18\pm0.02$ ($\alpha=3p/4-1/4=1.68\pm0.02$).
After the plateau, depending on the light curve modelling, the measured decay index
is either $\alpha=\alpha_{3}=1.08\pm0.02$ (Sect.~\ref{sec:beu3}; Fig.~\ref{f:fluxall_beu3}) or 
$\alpha=\alpha_{2}=2.0\pm0.8$ (Sect.~\ref{sec:beulore}; Fig.~\ref{f:fluxall_beulore}).
While the multiple smoothly broken power-law description (Sect.~\ref{sec:beu3}) definitely rules
out the wind environment, both environments are still possible in the two-component model
(Sect.~\ref{sec:beulore}), mainly because of the poorly measured decay index, $\alpha_2$.
If one interprets the flat decay as due to energy injection \citep{Nousek06,Zhang06}, the
corresponding index would be $q\sim0.3$.

In the off-axis jet interpretation, even if we consider an initially uniform sharp-edged jet,
the shocked external medium at the sides of the jet has a significantly smaller Lorentz factor
than near the head of the jet, and therefore its emission is not strongly beamed away from
off-beam lines of sight. As a result, either an early very shallow rise or decay is expected
for a realistic jet structure and dynamics \citep{Eichler06}.
In the case of XRF~030723, \citet{Granot05_2} showed that, for $\theta_{\rm obs}\sim\,2\,\theta_0$,
an initial plateau is expected in the light curve.

Our observations of the afterglow rise of XRF~080330 rule out the interpretation
proposed by \citet{Yamazaki09} of the plateau as due to an artifact of the choice of the
reference time, as all the other rising curves do.

\subsection{Jet break}
\label{sec:jetbreak}
According to the light curve description of Sect.~\ref{sec:beu3} shown in Fig.~\ref{f:fluxall_beu3},
for which only an ISM environment is possible (Sect.~\ref{sec:plateau}), after the plateau phase
the light curve is expected to approach the on-axis light curve with $\alpha=3(p-1)/4$.
The late-time steepening observed around $8\times10^4$~s, estimated to be
$\Delta\alpha=\alpha_4-\alpha_3=2.4\pm0.4$, cannot be produced by the passage of the
cooling frequency $\nu_{\rm c}$ through the optical, as that is expected to be as small as
$\Delta\alpha=1/4$ (ISM/wind).

In the off-axis jet interpretation, assuming a value for $\theta_0$ of a few degrees,
another advantage of this interpretation is the steep late time decay
(at $\gtrsim$~1~day) as a consequence of joining the post jet break on-axis light curve.
According to the light curve modelling given in Sect.~\ref{sec:beulore} shown in
Fig.~\ref{f:fluxall_beulore}, the gradual steepening following the plateau corresponds
to the post-jet break emission:
the observed power-law decay index, $\alpha_2=2.0\pm0.8$ is consistent with the
expected $\alpha$~$=$~$p$~$=$~$2.6$. We note that the relatively sharp jet break
favours the ISM environment.
Overall, in the context of an off-axis viewing angle interpretation, the light curve
suggests that $\theta_{\rm obs} \sim (1.5-2)\theta_0$ as well as an early jet break
(at $\lesssim\,1\;$day), which in turn implies a narrow jet with a half-opening angle
of the order of a few degrees, $\theta_0\,\sim\,0.05$. As a simple feasibility check,
we note that for $\theta_{\rm obs} < 2\theta_0$ the ratio of the on-axis to off-axis
$E_{\gamma,{\rm iso}}$ is equal to $\delta^2$ (assuming that the observed energy range includes
$E_{\rm p}$ where most of the energy is radiated), where $\delta$ is the ratio of their
corresponding Doppler factors and therefore of their $E_{\rm p}$ \citep{Eichler04}.
In our case, for an observed off-axis $(1+z)E_{\rm p} \sim 60\;$keV, an on-axis value of
$\sim 1\;$MeV would require $\delta = 1+[\Gamma_0(\theta_{\rm obs}-\theta_0)]^2 \sim 17$ and
$\Gamma_0(\theta_{\rm obs}-\theta_0) \sim 4$, which for $\theta_{\rm obs}-\theta_0 \sim (0.5-1)\theta_0$
and $\theta_0 \sim 0.05$ gives $\Gamma_0 \sim 80-160$. Here $\Gamma_0$ is the initial Lorentz factor
at the edge of the jet. More realistically, the jet would not be perfectly uniform with extremely
sharp edges, and instead $\Gamma_0$ is expected to be lower at the outer edge of the jet and larger
near its center (where it could easily reach several hundreds in our illustrative example here).
In this case $\delta^2\sim300$ so that the observed $E_{\gamma,{\rm iso}}$ in the $15-150\;$keV range,
which is $2\times 10^{51}\;$erg, would imply an on-axis value for $E_{\rm\gamma,iso}$ of $\sim10^{54}\;$erg,
which for a narrow jet with $\theta_{\rm obs} \sim 0.05$ would correspond to a true energy of the order
of $10^{51}\;$erg. This demonstrates that this scenario can work for reasonable values of the physical
parameters. We point out that the estimate of the on-axis $E_{\gamma,{\rm iso}}$ of $\sim10^{54}$~erg is
for a wide energy range containing $E_{\rm p}$, since in our illustrative example most of the energy
is released within the observed range.
A more accurate estimate of the break time and of the corresponding opening angle is difficult,
due to the degeneracy involved in the light curve modelling (Figs.~\ref{f:fluxall_beu3} and
\ref{f:fluxall_beulore}). Table~\ref{t:main_prop} summarises the main properties of XRF~080330.

\begin{table}
\caption{Summary of the main properties of XRF~080330.}
 \label{t:main_prop}
 \begin{tabular}{lc}
 \hline
 \hline
 \noalign{\smallskip}
Name    & Value\\
        &      \\
\noalign{\smallskip}
\hline
\noalign{\smallskip}
$z$         & $1.51$\\
$S(15-150~{\rm keV})$ & $(3.6\pm0.8)\times10^{-7}$~erg~cm$^{-2}$\\
$P(15-150~{\rm keV})$ & $(1.0\pm0.2)$~ph~cm$^{-2}$~s$^{-1}$\\
$E_{\rm p}$  & $<35$~keV\\
$E_{\rm p,i}=E_{\rm p}\, (1+z)$ & $<88$~keV\\
$E_{\rm iso}$ ($15$--$150$~keV, obs frame) & $(2.1\pm0.5)\times10^{51}$ ergs\\
$E_{\rm iso}$ ($1$--$10^4$~keV, GRB frame) & $<2.2\times10^{52}$ ergs\\
$t_{\rm jet}$ (obs frame) & $\lesssim1$~day\\
$\theta_0$ (jet opening angle)$^{\rm (a)}$ & few degrees\\
$\theta_{\rm obs}$ (viewing angle)$^{\rm (a)}$ & $\sim (1.5-2)\theta_0$\\
\noalign{\smallskip}
 \hline
 \end{tabular}
\flushleft
$^{\rm (a)}$ Under the assumption of a single off-axis jet.
\end{table}

\subsection{Late time red bump}
\label{sec:bump}
Overall, the two-component description of the light curves of Sect.~\ref{sec:beulore}
shown in Fig.~\ref{f:fluxall_beulore} appears to be slightly favoured over the multiple
smoothly broken power-law of Fig.~\ref{f:fluxall_beu3}.
So, irrespective of the nature of the rise and plateau, we speculate on the possible nature
of the second component, modelled in Fig.~\ref{f:fluxall_beulore} as a late time bump.
Clearly, a SN bump, such as that possibly observed in the light
curve of XRF~030723 \citep{Fynbo04}, is ruled out mainly because that is expected to peak
days later, which is incompatible with one single day after XRF~080330;
not to mention the too high redshift of XRF~080330 for a 1998bw-like SN to be detected.

Alternatively, a density bump seems a viable solution, given that $\nu_{\rm o}<\nu_{\rm c}$
(e.g. Lazzati et~al. 2002; Guidorzi et~al. 2005)\nocite{Lazzati02,Guidorzi05}, although
the explanation of the observed contemporaneous reddening requires ad hoc assumptions,
such as the case of GRB~050721, which showed similar properties to XRF~080330 (same
$\beta_{\rm ox}$ with no breaks between optical and X-ray, late time redder optical bump).
In that case, the observed reddening was explained as due to the presence of very dense
clumps surviving the GRB radiation and with a small covering factor \citep{Antonelli06}.

If the late reddening is due to the passage of $\nu_{\rm c}$ through the optical bands,
in addition to what argued in Sect.~\ref{sec:jetbreak}, another weak point of the multiple
smoothly broken power-law description of Fig.~\ref{f:fluxall_beu3} (Sect.~\ref{sec:beu3}) is the
chromatic change of $\Delta\beta_{\rm o}=0.26\pm0.06$ we observe in the optical bands around
$8\times10^4$~s. The passage of the cooling frequency does not explain it:
the observed reddening would be 0.5, i.e. twice as much. This could still be
the case, if our measurement might have taken place in the course of the spectral change,
as the broken power-law spectrum is a simple approximation.
However, since it is $\nu_{\rm c}>\nu_{\rm x}$ during the plateau because of the unbroken power-law
spectrum between optical and X-ray, $\nu_{\rm c}$ would have decreased very rapidly,
thus making this option not reasonable: if at $10^3$~s it is $\nu_{\rm c}>\nu_{\rm x}=10^{18}$~Hz,
at $10^5$~s it should be $\nu_{\rm c}>10^{17}$~Hz~$\gg\,\nu_{\rm o}$, because
$\nu_{\rm c}\propto\,t^{-1/2}$ (ISM case), thus this possibility is to be ruled out.
Likewise, in the wind case it is $\nu_{\rm c}\propto\,t^{1/2}$. The times $t_{K}$ and $t_{g}$,
at which $\nu_{\rm c}$ would cross the most redward and blueward filters, $K$ and $g'$, would
differ by a factor of $(\nu_g/\nu_K)^2\sim20$, which looks incompatible with the light curves.
Furthermore, the observed reddening rules out the wind case, as $\beta_{\rm o}$ should decrease
from $p/2$ to $(p-1)/2$.

Although an energy injection to the blast wave (forward shock) can explain the bump feature in
the light curve, it is difficult to explain the reddening if we consider only the forward shock
emission, as we have discussed.
A possible explanation for the reddening is that the rebrightening is due to the short-lived
($\Delta\,t\sim\,t$) reverse shock of a slow shell which caught up with the shock front
and increased its energy through a refreshed shock (e.g. Kumar \& Piran 2000; Granot
et~al. 2003; J\'ohannesson et~al. 2006):\nocite{KumarPiran00,Granot03,Johannesson06}
since that shock is going into a shell of ejecta, rather than the external medium,
it can have a much larger $\epsilon_B$ (magnetised fireball: Zhang et al. 2003;
Kumar \& Panaitescu 2003; Gomboc et~al. 2008b)\nocite{Zhang03,Kumar03,Gomboc08b}
and, therefore, a lower $\nu_{\rm c}$, quite naturally;
for an ISM where $\nu_{\rm c}$ decreases with time, then $\nu_{\rm c}$
of the reverse shock could be around the optical for reasonable model parameter values.
The need for a separate component to explain a chromatic break in the light curve was also
suggested in the case of GRB~061126 \citep{Gomboc08b}.
Notably, the final steepening after $10^5$~s, which in the modelling
of Sect.~\ref{sec:beu3} (Fig.~\ref{f:fluxall_beu3}) is described with $\alpha_4=3.5_{-0.3}^{+0.4}$,
is compatible with the high-latitude emission of the reverse shock: $\alpha=\beta_{\rm o,late}+2=3.05\pm0.06$.
The jet break might happen slightly earlier than the break time in the optical light curve.

A somewhat more contrived way to explain the bump is the appearance of a second narrower jet in the
two-component jet model, as proposed for XRF~030723 \citep{Huang04}. In this model, the viewing angle,
is within or slightly off the cone of the wide jet and outside the narrow jet.
The plateau phase would reflect the deceleration of the wide jet \citep{Granot06}.
Depending on the isotropic-equivalent kinetic energy of the wide and narrow jet, $E_{\rm w,iso}$
and $E_{\rm n,iso}$, on the jet opening angles, $\theta_{0,{\rm w}}$ and
$\theta_{0,{\rm n}}$, on the initial bulk Lorentz factors, $\gamma_{0,{\rm w}}$ and
$\gamma_{0,{\rm n}}$, respectively, as well as on the viewing angle, the afterglow emission
of either component is dominant at different times. According to the results of \citet{Huang04},
the light curve of XRF~080330 could be qualitatively explained as follows: the first component
obtained in Sect.~\ref{sec:beulore} (Fig.~\ref{f:fluxall_beulore}) represents the contribution
of the wide jet dominating at early times: the rise could be due either to the afterglow onset
(in a wind environment) or to a viewing angle slightly beyond the wide jet opening angle:
$\theta_{\rm obs}\gtrsim\theta_{0,{\rm w}}$. The appearance of the second component would mark
the deceleration and lateral expansion of the narrow jet, the peak time corresponding to
the case $\gamma_{\rm n}\sim1/\theta_{\rm obs}$.
Unlike XRF~030723, which showed a relatively sharp late-time peak, the bump exhibited by
XRF~080330 looks less sharp and pronounced. Although this might suggest a relatively lower
energy of the narrow jet compared to XRF~030723, yet we cannot exclude the case
$E_{\rm w,iso}\ll\,E_{\rm n,iso}$. The ratio of the observed energies of XRF~080330, of about $0.6$
according to the modelling of Sect.~\ref{sec:beulore}, corresponds to the ratio between the
early and the late time emissions of the wide and narrow jets, respectively.
Depending on the values of $\theta_{\rm obs}$, $\theta_{0,{\rm w}}$, $\theta_{0,{\rm n}}$,
a comparable ratio of observed energies, such as that observed for XRF~080330, can still be
obtained in the case of a much more energetic narrow jet, $E_{\rm w,iso}\ll\,E_{\rm n,iso}$.

Such a model turned out to be successful in accounting for the naked-eye GRB~080319B \citep{Racusin08}:
in that case, the two collimation-corrected energies were comparable, while the isotropic-equivalent
energy of the narrow jet ($\theta_{0,{\rm n}}=0.2^{\circ}$) was about 400 times larger
than that of the wide jet ($\theta_{0,{\rm w}}=4.0^{\circ}$).

However, in the context of the two-component model the explanation of the late time reddening
simultaneously with the appearance of the narrow component emission requires two different values
of $p$ for the two jets, which does not look reasonable on a physical ground.
Another option is that the cooling frequency of the second jet might be around the optical
at the time of the bump: however, if the shock micro-physics parameters ($p$, $\epsilon_B$,
$\epsilon_e$), are the same for the two shocks, as expected on physical grounds, and obviously
the external medium is the same, then the only thing that is different as far as $\nu_{\rm c}$
is concerned, is $E_{\rm iso}$. Since the presence of the bump requires $E_{\rm n,iso}>E_{\rm w,iso}$,
then this would not work for the wind case, where $\nu_{\rm c}\propto\,E_{\rm iso}^{1/2}$.
Even in the ISM case, where $\nu_{\rm c}\propto\,E_{\rm iso}^{-1/2}$, the fact that $\nu_{\rm c}>\nu_{\rm x}$
at $10^3$~s, and therefore $\nu_{\rm c}>10^{17}$~Hz at $10^5$~s for the wide jet,
requires $E_{\rm n,iso}/E_{\rm w,iso}\gtrsim10^4$ (which is very extreme) in order for
$\nu_{\rm c,n}(10^5~{\rm s})$ to be around $10^{15}$~Hz (required by the observed reddening).
Therefore, while this could work in principle, in practise it requires extreme parameters.
In particular, the amplitude of the bump suggests that $E_{\rm n}\sim\,E_{\rm w}$, and therefore
from the required $E_{\rm n,iso}/E_{\rm w,iso}\gtrsim10^4$ it would follow
$\theta_{\rm n}/\theta_{\rm w}\lesssim\,10^{-2}$, which seems pushed to the extreme.

Overall, these considerations make the single off-axis jet interpretation much more compelling.

\section{Conclusions}
\label{sec:conc}
XRF~080330 is representative of the XRR and XRF classes of soft GRBs. Its $\gamma$- and X-ray
properties of both prompt and high-energy afterglow emission place it in the low-flux
end of the distribution. The multi-band (NIR through UV) optical curve showed an initial rise
up to $\sim$300~s, followed by a $\sim$2-ks long plateau, temporally coinciding with the
canonical flat decay of X-ray afterglows of all kinds of GRBs, followed by a gradual steepening
and a possible jet break.
We provided two alternative descriptions of the light curve: a piecewise power-law with
three break times, the last of which occurring around $8\times10^4$~s and followed by a sharp
steepening, with the power-law decay index changing from $1.1$ to $3.5_{-0.3}^{+0.4}$.
The SED from NIR to X-ray wavelength is fitted with a simple power-law with $\beta_{\rm ox}=0.8$
and negligible GRB-frame extinction, $A_V<0.02$ adopting a SMC-like profile, with no evidence
for chromatic evolution during the rise, plateau and early ($<8\times10^4$~s) decay phases.
However, after the possible late time break we observe a reddening in the optical bands
of $\Delta\beta_{\rm o}=0.26\pm0.06$, which cannot be accounted for in terms of the synchrotron
spectrum evolution of a standard afterglow model, unless a different description of the light curve
is considered. In the alternative model of the light curves, we identified two distinct
components: the first is modelled with a smoothly broken power law and fits the rise plateau
and early decay of the afterglow, while the second, taking
over around $8\times10^4$~s, is modelled with an energy injection episode peaking
at $34_{-8}^{+11}$~ks and with a time-integrated energy of $\sim$60\% that of the first
component.

The X-ray light curve consists of the initial steep decay, which is likely the high-latitude
emission of the last $\gamma$-ray pulse. At the same time, the optical afterglow rises
up to a plateau, temporally coincident with the X-ray flat decay. In this case, we collected
strong evidence that the emission mechanism during this phase is the same from optical
to X-rays and is consistent with synchrotron emission of a decelerating fireball with
an electron energy distribution power-law index of $p=2.6$.

The lack of spectral evolution throughout the rise, plateau and early decay argue against
a temporally decreasing dust column density claimed to explain similar optical light
curves of past soft bursts.

The optical rise ($\alpha\sim-0.4$) is too slow for the afterglow
onset within a uniform circumburst medium, but could still be the case if a wind environment
is considered. In this case, under standard assumptions we constrained the Lorentz
factor of the fireball to be smaller than 80, thus confirming the scenario of XRFs as 
less relativistic GRBs. However, we found that the interpretation of a
single-component off-axis jet with an opening angle of a few degrees and a viewing angle
about twice as large, can explain the observations: this not only accounts
for the light curve morphology, but also explains the soft nature of XRF of
the $\gamma$-ray prompt event. The reddening observed at $8\times10^4$~s can be interpreted
as the short-lived reverse shock of an energy injection caused by a slow shell which caught
up with the fireball shock front, also responsible for the contemporaneous bump in the light curve.
A two-component jet could also work, but would introduce more free parameters and would require
extreme conditions.

The interpretation of the late bump as produced by a density enhancement in the medium swept
up by the fireball cannot be ruled out, although the reddening seems to require ad hoc
explanations. In this case, as shown by \citet{Nakar07}, it is hard to produce a flux
enhancement with density inhomogeneities, although it is not excluded given the lack of
a sharp rise in this bump.

Overall, the XRF~080330 optical and X-ray afterglows properties have also been observed
in many other GRBs \citep{Panaitescu08}. This both supports the view of a common
origin of XRFs and classical GRBs, which form a continuum and do not call for distinct
mechanisms. The importance of a prompt multi-wavelength coverage of the early
phases of a GRB is clearly demonstrated in the case of XRF~080330.


\acknowledgements{}

This work is supported by ASI grant I/R/039/04 and by the Ministry of University and Research of Italy
(PRIN 2005025417). JG gratefully acknowledges a Royal Society Wolfson Research Merit Award.
DARK is funded by the DNRF. PJ acknowledges support by a Marie Curie European Re-integration Grant
within the 7th European Community Framework Program under contract number PERG03-GA-2008-226653,
and a Grant of Excellence from the Icelandic Research Fund.
We gratefully acknowledge the contribution of the Swift team members at OAB, PSU, UL, GSFC, ASDC,
MSSL and our sub-contractors, who helped make this mission possible. We acknowledge Sami-Matias Niemi
for executing the NOT observations. CG is grateful to A.~Kann for his reading and comments.



\longtab{5}{
\begin{longtable}{rrrcr}
\caption{\label{tab:photom} Optical photometric set of XRF~080330.}\\
\hline\hline
Mid time         &     Exposure &   Magnitude$^{\mathrm{a,b}}$  &  Filter & Telescope   \\
      (s)        & (s)       	&        		  &         &  \\
\noalign{\smallskip}
\hline
\noalign{\smallskip}
\endfirsthead
\caption{continued.}\\
\hline\hline\hline
Mid Time         &     Exposure &   Mag$^{\mathrm{a,b}}$  &  Filter & Telescope   \\
      (s)        & (s)       	&        		  &         &  \\ 
\noalign{\smallskip}
\hline
\noalign{\smallskip}
\endhead
\hline
\endfoot
$       438$ & $    30$ & $17.68  \pm 0.03      $ & SDSS-$g$ & LT \\
$       695$ & $    30$ & $17.74  \pm 0.04      $ & SDSS-$g$ & LT \\
$       963$ & $    60$ & $17.75  \pm 0.03      $ & SDSS-$g$ & LT \\
$      1323$ & $    60$ & $17.82  \pm 0.04      $ & SDSS-$g$ & LT \\
$      1677$ & $    60$ & $17.98  \pm 0.03      $ & SDSS-$g$ & LT \\
$      2036$ & $    60$ & $18.07  \pm 0.03      $ & SDSS-$g$ & LT \\
$      2385$ & $    60$ & $18.27  \pm 0.05      $ & SDSS-$g$ & LT \\
$      2737$ & $    60$ & $18.36  \pm 0.04      $ & SDSS-$g$ & LT \\
$      3081$ & $    60$ & $18.50  \pm 0.05      $ & SDSS-$g$ & LT \\
$      3426$ & $    60$ & $18.57  \pm 0.05      $ & SDSS-$g$ & LT \\
$      3832$ & $   120$ & $18.76  \pm 0.03      $ & SDSS-$g$ & LT \\
\hline
$       186$ & $    10$ & $17.65  \pm 0.05      $ & SDSS-$r$ & LT \\
$       207$ & $    10$ & $17.60  \pm 0.03      $ & SDSS-$r$ & LT \\
$       228$ & $    10$ & $17.51  \pm 0.03      $ & SDSS-$r$ & LT \\
$       295$ & $    10$ & $17.39  \pm 0.05      $ & SDSS-$r$ & LT \\
$       317$ & $    10$ & $17.34  \pm 0.04      $ & SDSS-$r$ & LT \\
$       339$ & $    10$ & $17.37  \pm 0.03      $ & SDSS-$r$ & LT \\
$       609$ & $    30$ & $17.38  \pm 0.04      $ & SDSS-$r$ & LT \\
$       864$ & $    30$ & $17.43  \pm 0.04      $ & SDSS-$r$ & LT \\
$      1204$ & $    60$ & $17.44  \pm 0.05      $ & SDSS-$r$ & LT \\
$      1559$ & $    60$ & $17.58  \pm 0.03      $ & SDSS-$r$ & LT \\
$      1918$ & $    60$ & $17.72  \pm 0.04      $ & SDSS-$r$ & LT \\
$      2268$ & $    60$ & $17.86  \pm 0.04      $ & SDSS-$r$ & LT \\
$      2622$ & $    60$ & $17.98  \pm 0.03      $ & SDSS-$r$ & LT \\
$      2967$ & $    60$ & $18.10  \pm 0.03      $ & SDSS-$r$ & LT \\
$      3312$ & $    60$ & $18.22  \pm 0.05      $ & SDSS-$r$ & LT \\
$      3659$ & $    60$ & $18.35  \pm 0.03      $ & SDSS-$r$ & LT \\
$      4579$ &$3\times10$ & $18.65  \pm 0.04      $ & SDSS-$r$ & LT \\
$      4688$ & $    30$ & $18.65  \pm 0.04      $ & SDSS-$r$ & LT \\
$     30383$ & $   200$ & $20.78  \pm 0.06      $ & SDSS-$r$ & FTN \\
$     30605$ & $   200$ & $20.89  \pm 0.06      $ & SDSS-$r$ & FTN \\
$     30828$ & $   200$ & $20.90  \pm 0.06      $ & SDSS-$r$ & FTN \\
$     31050$ & $   200$ & $20.91  \pm 0.06      $ & SDSS-$r$ & FTN \\
$     31271$ & $   200$ & $20.92  \pm 0.06      $ & SDSS-$r$ & FTN \\
$     31494$ & $   200$ & $20.84  \pm 0.06      $ & SDSS-$r$ & FTN \\
$     31717$ & $   200$ & $20.89  \pm 0.06      $ & SDSS-$r$ & FTN \\
$     31940$ & $   200$ & $20.79  \pm 0.06      $ & SDSS-$r$ & FTN \\
$     32163$ & $   200$ & $20.86  \pm 0.07      $ & SDSS-$r$ & FTN \\
$     32385$ & $   200$ & $20.71  \pm 0.07      $ & SDSS-$r$ & FTN \\
$    116557$ &$20\times200$& $23.3   \pm 0.2       $ & SDSS-$r$ & FTN \\
\hline
$       523$ & $    30$ & $17.22  \pm 0.04      $ & SDSS-$i$ & LT \\
$       779$ & $    30$ & $17.20  \pm 0.05      $ & SDSS-$i$ & LT \\
$      1080$ & $    60$ & $17.26  \pm 0.04      $ & SDSS-$i$ & LT \\
$      1439$ & $    60$ & $17.38  \pm 0.03      $ & SDSS-$i$ & LT \\
$      1799$ & $    60$ & $17.50  \pm 0.04      $ & SDSS-$i$ & LT \\
$      2155$ & $    60$ & $17.64  \pm 0.05      $ & SDSS-$i$ & LT \\
$      2503$ & $    60$ & $17.75  \pm 0.03      $ & SDSS-$i$ & LT \\
$      2850$ & $    60$ & $17.88  \pm 0.04      $ & SDSS-$i$ & LT \\
$      3195$ & $    60$ & $17.98  \pm 0.05      $ & SDSS-$i$ & LT \\
$      3540$ & $    60$ & $18.12  \pm 0.05      $ & SDSS-$i$ & LT \\
$      4908$ & $    30$ & $18.56  \pm 0.05      $ & SDSS-$i$ & LT \\
$     64063$ &$4\times150$ & $21.27  \pm 0.09      $ & SDSS-$i$ & LT \\
$    120821$ &$10\times200$& $>21.0$                & SDSS-$i$ & FTN \\
\hline
$      2890$ & $    1476$ & $18.24 \pm  0.02$     & SDSS-$g$ & GROND \\
$      4699$ & $    1414$ & $18.88 \pm  0.02$     & SDSS-$g$ & GROND \\
$      6006$ & $     459$ & $19.20 \pm  0.02$     & SDSS-$g$ & GROND \\
$      6786$ & $     459$ & $19.40 \pm  0.02$     & SDSS-$g$ & GROND \\
$     85843$ & $   11522$ & $22.60 \pm  0.02$     & SDSS-$g$ & GROND \\
$    171989$ & $    5866$ & $>24.46$        & SDSS-$g$ & GROND \\
$    254165$ & $    5655$ & $>24.69$        & SDSS-$g$ & GROND \\
\hline
$      2890$ & $    1476$ & $18.05 \pm  0.02$     & SDSS-$r$ & GROND \\
$      4699$ & $    1414$ & $18.65 \pm  0.02$     & SDSS-$r$ & GROND \\
$      6006$ & $     459$ & $18.99 \pm  0.02$     & SDSS-$r$ & GROND \\
$      6786$ & $     459$ & $19.18 \pm  0.02$     & SDSS-$r$ & GROND \\
$     85843$ & $   11522$ & $22.29 \pm  0.02$     & SDSS-$r$ & GROND \\
$    171989$ & $    5866$ & $>24.15$     & SDSS-$r$ & GROND \\
$    254165$ & $    5655$ & $>24.33$     & SDSS-$r$ & GROND \\
\hline
$      2890$ & $    1476$ & $17.86 \pm  0.02$    & SDSS-$i$ & GROND \\
$      4699$ & $    1414$ & $18.43 \pm  0.02$    & SDSS-$i$ & GROND \\
$      6006$ & $     459$ & $18.77 \pm  0.02$    & SDSS-$i$ & GROND \\
$      6786$ & $     459$ & $18.95 \pm  0.02$    & SDSS-$i$ & GROND \\
$     85843$ & $   11522$ & $22.00 \pm  0.03$    & SDSS-$i$ & GROND \\
$    171989$ & $    5866$ & $>23.28$    & SDSS-$i$ & GROND \\
$    254165$ & $    5655$ & $>23.47$    & SDSS-$i$ & GROND \\
\hline
$      2890$ & $    1476$ & $17.69 \pm  0.02$    & SDSS-$z$ & GROND \\
$      4699$ & $    1414$ & $18.27 \pm  0.02$    & SDSS-$z$ & GROND \\
$      6006$ & $     459$ & $18.58 \pm  0.02$    & SDSS-$z$ & GROND \\
$      6786$ & $     459$ & $18.77 \pm  0.03$    & SDSS-$z$ & GROND \\
$     85843$ & $   11522$ & $21.82 \pm  0.04$    & SDSS-$z$ & GROND \\
$    171989$ & $    5866$ & $>22.68$    & SDSS-$z$ & GROND \\
$    254165$ & $    5655$ & $>23.16$    & SDSS-$z$ & GROND \\
\hline
$       228$ & $      60$ & $16.90 \pm  0.10$    & $J^{\mathrm{(c)}}$ & GROND \\
$       336$ & $      60$ & $16.92 \pm  0.10$    & $J^{\mathrm{(c)}}$ & GROND \\
$       559$ & $      60$ & $16.86 \pm  0.10$    & $J^{\mathrm{(c)}}$ & GROND \\
$       768$ & $      60$ & $16.88 \pm  0.10$    & $J^{\mathrm{(c)}}$ & GROND \\
$       871$ & $      60$ & $16.85 \pm  0.10$    & $J^{\mathrm{(c)}}$ & GROND \\
$       972$ & $      60$ & $16.92 \pm  0.10$    & $J^{\mathrm{(c)}}$ & GROND \\
$      1130$ & $     120$ & $16.93 \pm  0.10$    & $J^{\mathrm{(c)}}$ & GROND \\
$      1317$ & $     120$ & $17.04 \pm  0.10$    & $J^{\mathrm{(c)}}$ & GROND \\
$      1504$ & $     120$ & $17.00 \pm  0.10$    & $J^{\mathrm{(c)}}$ & GROND \\
$      2920$ & $    1200$ & $17.48 \pm  0.10$    & $J^{\mathrm{(c)}}$ & GROND \\
$      4736$ & $    1200$ & $18.02 \pm  0.10$    & $J^{\mathrm{(c)}}$ & GROND \\
$      6034$ & $     950$ & $18.34 \pm  0.10$    & $J^{\mathrm{(c)}}$ & GROND \\
$     85868$ & $    9960$ & $21.61 \pm  0.11$    & $J^{\mathrm{(c)}}$ & GROND \\
$    172013$ & $    5160$ & $>21.99$    & $J^{\mathrm{(c)}}$ & GROND \\
$    254200$ & $    4800$ & $>22.15$    & $J^{\mathrm{(c)}}$ & GROND \\
\hline
$       228$ & $      60$ & $16.78 \pm  0.10$    & $H^{\mathrm{(c)}}$ & GROND \\
$       336$ & $      60$ & $16.70 \pm  0.10$    & $H^{\mathrm{(c)}}$ & GROND \\
$       448$ & $      60$ & $16.65 \pm  0.10$    & $H^{\mathrm{(c)}}$ & GROND \\
$       559$ & $      60$ & $16.70 \pm  0.10$    & $H^{\mathrm{(c)}}$ & GROND \\
$       667$ & $      60$ & $16.62 \pm  0.10$    & $H^{\mathrm{(c)}}$ & GROND \\
$       768$ & $      60$ & $16.68 \pm  0.10$    & $H^{\mathrm{(c)}}$ & GROND \\
$       871$ & $      60$ & $16.65 \pm  0.10$    & $H^{\mathrm{(c)}}$ & GROND \\
$       972$ & $      60$ & $16.70 \pm  0.10$    & $H^{\mathrm{(c)}}$ & GROND \\
$      1130$ & $     120$ & $16.70 \pm  0.10$    & $H^{\mathrm{(c)}}$ & GROND \\
$      1317$ & $     120$ & $16.70 \pm  0.10$    & $H^{\mathrm{(c)}}$ & GROND \\
$      1504$ & $     120$ & $16.74 \pm  0.10$    & $H^{\mathrm{(c)}}$ & GROND \\
$      1690$ & $     120$ & $16.85 \pm  0.10$    & $H^{\mathrm{(c)}}$ & GROND \\
$      2920$ & $    1200$ & $17.24 \pm  0.10$    & $H^{\mathrm{(c)}}$ & GROND \\
$      4736$ & $    1200$ & $17.78 \pm  0.10$    & $H^{\mathrm{(c)}}$ & GROND \\
$      6034$ & $     950$ & $18.06 \pm  0.10$    & $H^{\mathrm{(c)}}$ & GROND \\
$     85868$ & $    9960$ & $21.16 \pm  0.11$    & $H^{\mathrm{(c)}}$ & GROND \\
$    172013$ & $    5160$ & $>21.36$    & $H^{\mathrm{(c)}}$ & GROND \\
$    254200$ & $    4800$ & $>21.28$    & $H^{\mathrm{(c)}}$ & GROND \\
\hline
$       228$ & $      60$ & $16.73 \pm  0.10$    & $K^{\mathrm{(c)}}$ & GROND \\
$       336$ & $      60$ & $16.47 \pm  0.10$    & $K^{\mathrm{(c)}}$ & GROND \\
$       448$ & $      60$ & $16.56 \pm  0.10$    & $K^{\mathrm{(c)}}$ & GROND \\
$       559$ & $      60$ & $16.42 \pm  0.10$    & $K^{\mathrm{(c)}}$ & GROND \\
$       667$ & $      60$ & $16.57 \pm  0.10$    & $K^{\mathrm{(c)}}$ & GROND \\
$       768$ & $      60$ & $16.46 \pm  0.10$    & $K^{\mathrm{(c)}}$ & GROND \\
$       871$ & $      60$ & $16.45 \pm  0.10$    & $K^{\mathrm{(c)}}$ & GROND \\
$       972$ & $      60$ & $16.55 \pm  0.10$    & $K^{\mathrm{(c)}}$ & GROND \\
$      1317$ & $     120$ & $16.45 \pm  0.10$    & $K^{\mathrm{(c)}}$ & GROND \\
$      1504$ & $     120$ & $16.53 \pm  0.10$    & $K^{\mathrm{(c)}}$ & GROND \\
$      1690$ & $     120$ & $16.77 \pm  0.10$    & $K^{\mathrm{(c)}}$ & GROND \\
$      2920$ & $    1200$ & $17.08 \pm  0.10$    & $K^{\mathrm{(c)}}$ & GROND \\
$      4736$ & $    1200$ & $17.62 \pm  0.10$    & $K^{\mathrm{(c)}}$ & GROND \\
$     85868$ & $    9960$ & $20.93 \pm  0.13$    & $K^{\mathrm{(c)}}$ & GROND \\
$    172013$ & $    5160$ & $>20.37$    & $K^{\mathrm{(c)}}$ & GROND \\
$    254200$ & $    4800$ & $>20.49$    & $K^{\mathrm{(c)}}$ & GROND \\
\hline
$      1238$ & $      60$ & $17.33 \pm 0.02 $    & $R$ & NOT \\
$      1609$ & $     300$ & $17.46 \pm 0.01 $    & $R$ & NOT \\
$      1927$ & $      60$ & $17.61 \pm 0.01 $    & $R$ & NOT \\
\hline
$      89.4$ & $    30$ & $17.90  \pm 0.19      $ & SDSS-$r$ equivalent & REM \\
$     167.6$ & $    30$ & $17.61  \pm 0.13      $ & SDSS-$r$ equivalent & REM \\
$     246.2$ & $    30$ & $17.41  \pm 0.12      $ & SDSS-$r$ equivalent & REM \\
$     354.2$ & $    30$ & $17.34  \pm 0.12      $ & SDSS-$r$ equivalent & REM \\
$     432.9$ & $    30$ & $17.40  \pm 0.11      $ & SDSS-$r$ equivalent & REM \\
$     511.1$ & $    30$ & $17.42  \pm 0.13      $ & SDSS-$r$ equivalent & REM \\
$     590.5$ & $    30$ & $17.41  \pm 0.13      $ & SDSS-$r$ equivalent & REM \\
$     663.6$ & $    30$ & $17.42  \pm 0.08      $ & SDSS-$r$ equivalent & REM \\
$     732.7$ & $    30$ & $17.48  \pm 0.09      $ & SDSS-$r$ equivalent & REM \\
$     802.7$ & $    30$ & $17.47  \pm 0.10      $ & SDSS-$r$ equivalent & REM \\
$     871.8$ & $    30$ & $17.40  \pm 0.09      $ & SDSS-$r$ equivalent & REM \\
$     971.1$ & $    60$ & $17.45  \pm 0.05      $ & SDSS-$r$ equivalent & REM \\
$    1099.9$ & $    60$ & $17.50  \pm 0.06      $ & SDSS-$r$ equivalent & REM \\
$    1403.1$ & $    30$ & $17.65  \pm 0.11      $ & SDSS-$r$ equivalent & REM \\
$    1471.4$ & $    30$ & $17.57  \pm 0.10      $ & SDSS-$r$ equivalent & REM \\
$    1540.5$ & $    30$ & $17.62  \pm 0.11      $ & SDSS-$r$ equivalent & REM \\
$    1640.7$ & $    60$ & $17.58  \pm 0.06      $ & SDSS-$r$ equivalent & REM \\
$    1769.5$ & $    60$ & $17.73  \pm 0.07      $ & SDSS-$r$ equivalent & REM \\
\hline
$      50.4$ & $  60.0$ & $> 16.6    $ & clear & TAROT \\
$     120.9$ & $  67.8$ & $17.62  \pm 0.60      $ & clear & TAROT \\
$     194.1$ & $  66.6$ & $17.38  \pm 0.46      $ & clear & TAROT \\
$     249.3$ & $  30.6$ & $16.96  \pm 0.40      $ & clear & TAROT \\
$     332.7$ & $  89.4$ & $17.17  \pm 0.25      $ & clear & TAROT \\
$     429.0$ & $  91.2$ & $17.07  \pm 0.20      $ & clear & TAROT \\
$     528.7$ & $  90.0$ & $16.83  \pm 0.29      $ & $R$   & TAROT \\
$     626.4$ & $  90.0$ & $16.83  \pm 0.22      $ & clear & TAROT \\
 $     723.0$ & $  90.0$ & $16.86  \pm 0.22      $ & clear & TAROT \\ 
$     927.0$ & $  90.0$ & $17.52  \pm 0.46      $ & clear & TAROT \\
$    1023.9$ & $  89.4$ & $17.15  \pm 0.30      $ & clear & TAROT \\
$    1123.9$ & $  90.0$ & $17.00  \pm 0.39      $ & $R$   & TAROT \\
$    1222.5$ & $  89.4$ & $16.73  \pm 0.21      $ & clear & TAROT \\
$    1318.5$ & $  89.4$ & $16.75  \pm 0.25      $ & clear & TAROT \\
$    1418.3$ & $  90.0$ & $>16.8                $ & $R$   & TAROT \\
\hline
$        68$ & $    10$ & $>17.38     $ &$V$    & UVOT \\
$       103$ & $    40$ & $20.06  \pm 0.397     $ & white & UVOT \\
$       128$ & $    10$ & $18.55  \pm 0.20      $ & white & UVOT \\
$       138$ & $    10$ & $18.36  \pm 0.44      $ & white & UVOT \\
$       148$ & $    10$ & $18.28  \pm 0.39      $ & white & UVOT \\
$       158$ & $    10$ & $18.39  \pm 0.43      $ & white & UVOT \\
$       168$ & $    10$ & $18.22  \pm 0.38      $ & white & UVOT \\
$       178$ & $    10$ & $18.05  \pm 0.32      $ & white & UVOT \\
$       214$ & $    50$ & $17.74  \pm 0.19      $ &$V$    & UVOT \\
$       264$ & $    50$ & $17.72  \pm 0.22      $ &$V$    & UVOT \\
$       314$ & $    50$ & $17.75  \pm 0.26      $ &$V$    & UVOT \\
$       364$ & $    50$ & $17.52  \pm 0.23      $ &$V$    & UVOT \\
$       414$ & $    50$ & $17.37  \pm 0.17      $ &$V$    & UVOT \\
$       464$ & $    50$ & $17.79  \pm 0.24      $ &$V$    & UVOT \\
$       514$ & $    50$ & $17.44  \pm 0.18      $ &$V$    & UVOT \\
$       557$ & $    36$ & $17.37  \pm 0.20      $ &$V$    & UVOT \\
$       629$ & $    20$ & $>18.02     $ & UVW1  & UVOT \\
$       654$ & $    20$ & $17.14  \pm 0.16      $ &$U$    & UVOT \\
$       673$ & $    10$ & $17.81  \pm 0.23      $ &$B$    & UVOT \\
$       688$ & $    10$ & $17.76  \pm 0.14      $ & white & UVOT \\
$       733$ & $    20$ & $17.74  \pm 0.34      $ &$V$    & UVOT \\
$       782$ & $    20$ & $> 18.04     $ & UVW1  & UVOT \\
$       788$ & $   172$ & $> 18.77     $ & UVW2  & UVOT \\
$       807$ & $    20$ & $16.66  \pm 0.17      $ &$U$    & UVOT \\
$       827$ & $    10$ & $17.66  \pm 0.22      $ &$B$    & UVOT \\
$       838$ & $    10$ & $17.55  \pm 0.12      $ & white & UVOT \\
$       848$ & $    10$ & $17.91  \pm 0.15      $ & white & UVOT \\
$       992$ & $   796$ & $> 18.55     $ & UVM2  & UVOT \\
$      1010$ & $    50$ & $17.37  \pm 0.17      $ &$V$    & UVOT \\
$      1060$ & $    50$ & $17.37  \pm 0.17      $ &$V$    & UVOT \\
$      1110$ & $    50$ & $17.41  \pm 0.17      $ &$V$    & UVOT \\
$      1160$ & $    50$ & $17.72  \pm 0.21      $ &$V$    & UVOT \\
$      1210$ & $    50$ & $17.36  \pm 0.21      $ &$V$    & UVOT \\
$      1260$ & $    50$ & $17.75  \pm 0.26      $ &$V$    & UVOT \\
$      1310$ & $    50$ & $18.11  \pm 0.29      $ &$V$    & UVOT \\
$      1360$ & $    51$ & $17.77  \pm 0.22      $ &$V$    & UVOT \\
$      1426$ & $    10$ & $17.67  \pm 0.29      $ & UVW1  & UVOT \\
$      1475$ & $    20$ & $18.14  \pm 0.22      $ &$B$    & UVOT \\
$     72100$ & $  5887$ & $> 20.10     $ &$V$    & UVOT \\
$    307644$ & $ 47917$ & $> 21.49     $ &$B$    & UVOT \\
$    307940$ & $ 47715$ & $> 20.53     $ &$V$    & UVOT \\
$    307583$ & $ 47956$ & $> 21.15     $ &$U$    & UVOT \\
$    307483$ & $ 48074$ & $> 21.58     $ & UVW1  & UVOT \\
$    391039$ & $ 40970$ & $> 23.32     $ & white & UVOT \\
\end{longtable}
  \begin{list}{}{} 
  \item[$^{\mathrm{a}}$] Values are not corrected for Galactic extinction.
  \item[$^{\mathrm{b}}$] Errors at the 68\% confidence level and
                  upper limits (3~$\sigma$) are given. 
  \item[$^{\mathrm{c}}$] AB magnitudes.
  \end{list}   
}

\begin{table*}
\caption{Lines detected in the XRF~080330 spectrum.}
\label{t:spectrum}
\centering
\begin{tabular}{r r l l c }
\hline\hline
$\lambda_{\textrm{obs}}$ [\AA] & $\lambda_{\textrm{rest}}$ [\AA] & \hspace*{2mm} $z$ & Feature &
EW$_{\textrm{obs}}$ [\AA]\\
\noalign{\smallskip}
\hline
\noalign{\smallskip}
3889.4 & 1548.2/1550.8 & 1.5101 & \mbox{C\,{\sc iv}}    & 4.3$\pm$0.7 \\
4028.8 & 1608.5        & 1.5047 &  FeII   & 2.0$\pm$0.7 \\
4198.2 & 1670.8        & 1.5127 &  AlII   & 2.0$\pm$0.7 \\
5093.5 & 2026.1        & 1.5139 & \mbox{Zn\,{\sc ii}} & 4.3$\pm$0.4 \\
5190.3 & 2062.2        & 1.5166 &  CrII   & \multirow{2}{*}{0.6$\pm$0.3} \\
5190.3 & 2062.7        & 1.5166 &  ZnII   &  \\
5883.4 & 2344.2        & 1.5100 &  FeII   & 1.4$\pm$0.5 \\
5983.0 & 2382.8        & 1.5109 &  FeII   & 2.2$\pm$0.5 \\
6497.5 & 2586.7        & 1.51   &  FeII & \multirow{2}{*}{11$\pm$2} \\
6527.0 & 2600.2        & 1.51   &  FeII &  \\
7031.1 & 2796.3/2803.5 & 1.5112 &  MgII   & 5.2$\pm$0.3 \\
\hline
\end{tabular}
\end{table*}

\end{document}